\documentclass[structabstract]{aa}  
\usepackage{graphicx}
\usepackage{longtable}
\usepackage{txfonts}
\begin{document}
\title{The Arches cluster revisited: I. Data presentation and stellar census  
\thanks{Based on observations made at the European Southern Observatory, 
Paranal, Chile under programs ESO 087.D-0317, 091.D-0187 and 099.D-0345.} 
}
\author{J.~S.~Clark\inst{1}
\and M.E.~Lohr\inst{1}
\and F.~Najarro\inst{2}
\and H.~Dong\inst{3}
\and F.~Martins\inst{4}}
\institute{
$^1$School of Physical Sciences, The Open 
University, Walton Hall, Milton Keynes, MK7 6AA, United Kingdom\\
$^2$Departamento de Astrof\'{\i}sica, Centro de Astrobiolog\'{\i}a, 
(CSIC-INTA), Ctra. Torrej\'on a Ajalvir, km 4,  28850 Torrej\'on de 
Ardoz, 
Madrid, Spain\\
$^3$Instituto de Astrof\'{i}sica de Andaluc\'{i}a (CSIC), Glorieta de la
Astronom\'{a} S/N, E-18008 Granada, Spain\\
$^4$LUPM, Universit\'{e} de Montpellier, CNRS, Place Eug\`{e}ne Bataillon, F-34095 
Montpellier, France
}

   \abstract{Located within the central region of the Galaxy, the Arches cluster appears to be one of the youngest, 
densest and most massive stellar aggregates within the Milky Way. As such it has the potential to be a uniquely 
instructive 
laboratory for the study of star formation in extreme environments and the physics of very massive stars.}
{To realise this possibility, the fundamental physical properties of both cluster and constituent stars need to be 
robustly determined; tasks we attempt here.}
{In order to accomplish these goals we provide and analyse new multi-epoch near-IR spectroscopic data obtained with the  
VLT/SINFONI and  photometry from the HST/WFC3. We are able to stack multiple epochs of spectroscopy for individual stars 
in order to obtain the deepest view of the cluster members ever obtained.}
{We present spectral classifications for 88 cluster  members, all of which are WNLh or O stars: a factor of three 
increase over  previous studies.  We find no further examples of  Wolf-Rayet stars within the cluster;
importantly no H-free examples were identified. The  smooth and continuous progression in spectral 
morphologies from O super-/hypergiants through to the WNLh cohort implies a direct evolutionary connection. We 
identify candidate giant and main sequence O stars spectroscopically for the first time. No products of 
binary evolution may be unambiguously identified despite the presence of massive binaries within the Arches.}
{Notwithstanding difficulties imposed by the highly uncertain (differential) reddening to the Arches, we infer a  main 
sequence/luminosity class V
turn-off mass of $\sim30-38M_{\odot}$ via the distribution of spectral types. Analysis of the eclipsing binary F2 
suggests  current masses of $\sim80M_{\odot}$ and $\sim60M_{\odot}$ for the WNLh and O hypergiant cohorts, 
respectively; we  conclude that all classified stars have masses $>20M_{\odot}$. An age of $\sim2.0-3.3$Myr is suggested 
by the   turn-off between $\sim$O4-5 V; constraints imposed by the 
supergiant population and the lack   of H-free WRs are  consistent with this estimate. While the absence of highly 
evolved WC stars strongly argues against the prior occurrence of SNe 
within the Arches, the derived age does accommodate such events for exceptionally massive stars. Further progress will require quantitative analysis of multiple individual 
cluster members in addition to further spectroscopic observations to better constrain the binary and main sequence 
populations; nevertheless it is abundantly clear that the Arches offers an unprecedented insight into 
the formation, evolution and death of the most massive stars Nature allows to form.
 } 

\maketitle

\section{Introduction}

Determining the  formation mechanism, properties and lifecycle 
of very massive stars is one of the most important unresolved issues
in stellar astrophysics; a problem exacerbated by their impact on galactic evolution 
- via radiative, mechanical and chemical feedback - and their role as 
  progenitors of some of the most luminous electromagnetic and gravitational wave
transients in the Universe. Currently, 
 even the most basic questions - such as how massive Nature permits stars to 
grow -  remain unanswered. How do they reach their final masses - do they 
form via a scaled up version of the disc-mediated accretion paradigm for 
low mass stars (Shu et al. \cite{shu}), competitive accretion in a 
clustered environment (Bonnell et al. \cite{bonnell}) or are they instead 
built-up by a more exotic avenue such as mergers, either before or during 
core-H burning (Schneider et al. \cite{schneider14}, \cite{schneider15})? Competitive accretion 
suggest that massive stars  should  form in stellar aggregates (clusters 
or OB associations) but is this always the case? And a related question - 
does the environment in which they form influence their final properties, 
such as occurrence of binarity and the form of the initial mass function 
(IMF)?

The central regions of our Galaxy provide a unique laboratory for the 
study of massive stars, hosting 3 young ($<10$ Myr), massive 
($\gtrsim10^4M_{\odot}$) clusters; the Galactic Centre, Quintuplet and the 
Arches. Critically, their co-location and consequently well defined distance aids
 luminosity determinations for cluster members; 
observationally 
challenging for isolated field stars in the galactic disc. Moreover the cluster ages they span 
means that evolutionary pathways for a wide range of initial masses
($\sim20-100M_{\odot}$) may be constrained via study of their evolved stellar populations.

We may also invert this argument, utilising these clusters to explore the effects 
of the extreme Galactic Centre environment on (massive) star formation and the role 
their subsequent feedback plays in the wider ecology of the circumnuclear molecular and 
starburst region. In this respect the proximity of the Galactic Centre renders it the 
sole testbed for studying the physics, recent star formation history and assemblage of 
the circumnuclear starburst of a galaxy at the level of individual constituent stars.

Of the three circumnuclear clusters, the Arches  (Cotera et al. \cite{cotera}) is of 
 particular interest, given  its (i) apparent youth and high mass - resulting in the 
 upper reaches of the IMF being well populated - and (ii) compactness - leading to  an 
extreme stellar density   (Figer et al. \cite{figer99b}, Martins et al. \cite{martins08}
 (henceforth Ma08)). Consequently numerous studies of the Arches have been 
 undertaken in the two decades following its discovery to better understand its bulk 
properties and those of its constituent stars. The majority of these have focused on
 determining the shape of the IMF and the presence of possible mass segregation, although
 such efforts  are hampered by uncertainties in the extinction law and significant 
differential reddening across the field (Figer et al. \cite{figer02} (henceforth Fi02), 
Stolte et al. \cite{stolte02}, \cite{stolte05}, Kim et al. \cite{kim}, Espinoza et al. \cite{espinoza}, 
Clarkson et al. \cite{clarkson} and Habibi et al. \cite{habibi}). 

Corresponding spectroscopic observations to classify and date the constituent stars and 
cluster are more limited (Fi02, Najarro et al. \cite{najarro}, Ma08) but reveal homogeneous populations of highly luminous 
WN7-9h and mid-O supergiants, with two mid-O hypergiants with spectral morphologies 
intermediate between these groups. Assuming uniform reddening,  non-LTE model-atmosphere 
analysis of these stars by Ma08 suggested that stars in both cohorts are 
very massive ($>60M_{\odot}$) 
and young, with a global age of 2-4 Myr suggested for the Arches. Intriguingly,
at the lower extent of the age-range no supernovae (SNe) would be expected to have occurred and 
hence the most massive stars born within the Arches should still be present  ($\gtrsim100M_{\odot}$;
 Crowther et al. \cite{crowther10}), allowing us to probe the upper reaches of the IMF and 
the stars that populate it. Moreover, the modelling results also hint at non-coevality for the cluster, with 
the more luminous WN7-9h stars potentially being younger than the less evolved supergiants 

Following the recognition of the prevalence and importance of binarity to understanding massive
 stellar evolution (Sana et al. \cite{sana12}, de Mink et al. \cite{demink}) Schneider et al. 
(\cite{schneider14}) re-interpreted the cluster (I)MF and age of the Arches under the assumption 
that {\em all} the massive stars within were binaries. As a result they revised the cluster age
 to a unique value of $3.5\pm0.7$ Myr, and concluded that the brightest $9.2\pm3$ stars (all 
WN7-9h) were the rejuvenated products of binary interaction - essentially indicating that very 
massive stars could form via a two-stage process with a second episode of mass-accretion onto the secondary
during the H-burning phase of the primary.

Given the extreme requirements placed on the binary population of the Arches by Schneider 
et al. (\cite{schneider14}) in order to explain the apparent non-coevality of the Arches 
(cf. Ma08), improved observational constraints on  the cluster properties are imperative.
 Moreover, due to the anticipated youth of the Arches, such studies would also provide 
important insights into the formation and lifecycle of extremely massive stars.  
Unfortunately, no systematic survey for binarity within the Arches  has yet
 been attempted, although indirect diagnostics suggest binarity  for a number of cluster 
members (e.g. Wang et al. \cite{wang}). A better determination of the cluster age would 
also require improved bolometric luminosity estimates - via individually determined corrections
 for interstellar reddening (Sect. 4) - in order to facilitate improved isochrone fitting, in 
conjunction with the identification of the main sequence turn-off.

In order to address these issue we undertook a multi-epoch spectroscopic survey of
 the Arches with the integral field spectrograph SINFONI mounted on the Very Large 
Telescope (VLT). These observations  permitted a search for both radial velocity and 
line profile variability, potentially indicative of reflex binary  motion and  wind 
collision zones, respectively. Moreover, stacking multiple individual observations 
allowed us to obtain higher signal/noise (S/N) spectra and hence reach less evolved 
cluster members than previous studies. These data were supplemented with new Hubble 
Space Telescope (HST) Wide Field Camera 3 (WFC3) photometry. In this work we present
 the resultant datasets and utilise them to provide an updated stellar census of the
 Arches and discuss the impact of the new data on the determination of cluster 
properties. Companion  papers provide a tailored quantitative analysis of the binary 
system F2 (adopting the naming convention of Fi02) and discuss the spectral variability of cluster 
members (Lohr et al. submitted and in prep.; henceforth papers II and III). 

This paper is structured as follows. In Sect. 2 we present the data acquisition and 
reduction strategies employed. In Sect. 3  we  provide 
 spectral classifications for cluster members, which are summarised in Table A.1. The
 important and complicating issue of interstellar extinction towards the Arches is 
investigated in Sect. 4, while in Sect. 5 we address the implications of our new results
 for the stellar masses of the cluster members and the determination of the cluster age.
Finally we discuss evolutionary implications in Sect. 6 and in Sect. 7 summarise our findings
 and highlight future prospects. 

\section{Data acquisition and reduction}

\subsection{Spectroscopy}

As initially envisaged, our programme was to closely follow the 
observational strategy of Ma08. Specifically, between April to August 2011
the SINFONI integral field spectrograph 
on the ESO/VLT (Eisenhauer et al. \cite{eisenhauer}; Bonnet et al. \cite{bonnet}) 
was used in service mode to make multiple K band observations of  three overlapping 
fields in the central Arches cluster, and seven fields on the periphery of the 
cluster. For each observation 
four 60~s exposures nodding across the target field were taken, interspersed with sky frames to
optimise background removal for the combined images.  Telluric
standards with spectral types B2--B9 were observed before or after
each set of science frames. 

However poor weather
significantly impacted on observations, leaving  the programme substantially
unfinished. As a consequence further time was sought and awarded, with observations
made in March to August 2013 and again in April to July 2017.
Unfortunately,  on each occasion the programmes remained incomplete at the end  of the semester. This 
resulted in a highly inhomogeneous composite dataset, with some fields visited multiple times,
 while other outliers were only observed once or twice. Moreover, the signal-to-noise (S/N)
 of individual integrations was highly variable given the poor observing conditions in which some observations 
were attempted. In order to generate the most complete dataset possible, previous
spectroscopic observations of the Arches utilising the same experimental set-up
were extracted from the archive\footnote{ESO proposals O87.D-0342 and 093.D-0306}, to be reduced in an identical manner (see 
below). A  further epoch of  spectroscopy covering outlying fields 
was extracted from data cubes used for Ma08\footnote{ESO proposal 075.D-0736}; in this 
instance  sky subtraction and telluric removal was carried out via the methodology  described in that paper.

A detailed breakdown of the timings of individual observations is provided in paper III
 and we refer the interested reader to this work. Foreshadowing 
the following discussion, since we are only interested in obtaining the highest 
quality summed spectra for this work we simply list the number of individual, 
contributing  observations for each cluster member in Table A.1. In a 
number of cases multiple individual observations were made on the same night - 
hence we also list the total number of nights (epochs) on which data were 
obtained. 

Science and telluric standard frames were reduced with the latest
version of the ESO SINFONI pipeline running under Reflex.  This
performed flat-fielding and optical distortion corrections, wavelength
calibration and improved sky background subtraction, before stacking 
slitlets into data cubes for each frame, and co-adding cubes for each
subfield and for each observation.  QFitsView was initially used to  
inspect the reduced
cubes\footnote{http://www.mpe.mpg.de/$\sim$ott/QfitsView/}, and then a
custom IDL code was written to facilitate manual extraction of spectra
for multiple individual pixels associated with each science target or
telluric standard star.  Care had to be taken to avoid selection of  
pixels contaminated by light from nearby objects; on certain epochs,
unwanted instrumental features were observable in specific regions of
the data cube, so these pixels were also excluded.

These pixel-spectra were then combined into a single spectrum per
object, using an approach based on the optimal extraction algorithm
for long-slit spectra of Horne (\cite{horne}).  Specifically, spatial
profiles were determined for each pixel-spectrum (indicating the
probability that a detected photon at a given wavelength would be
registered in that pixel) by dividing the pixel's flux at each
wavelength by the total flux over all pixels at that wavelength, and 
then median-smoothing the resulting profile estimates to reduce the
impact of bad lines in individual pixels.  Each pixel's fluxes were
then divided by its smoothed spatial profile to give an estimate of   
the total spectrum; the median of all such estimates was then taken as
our best estimate of the object's one-dimensional spectrum.

Preliminary radial velocity measurements indicated small but
significant errors in the wavelength solutions obtained by the
pipeline software.  Therefore, corrections were determined by
cross-correlating all science and telluric spectra with a standard
high resolution telluric spectrum in the $K$ band provided by
ESO\footnote{https://www.eso.org/sci/facilities/paranal/decommissioned/isaac/
  tools/spectra/atmos\_S\_K.fits}, using the IRAF \emph{telluric}  
task, and the headers adjusted accordingly.

The only intrinsic absorption line in our telluric spectra, in the
wavelength region of interest, was the Br$\gamma$ line at
2.166~$\mu$m.  This was removed by fitting it with a double Lorentzian
profile.  When two telluric standards had been observed in a given 
epoch, before and after a set of science frames, a custom telluric    
spectrum was created for each science observation by interpolating
between the two standards, to match the time of the science
observation, and normalising to the continuum.  Each science spectrum
was then divided by the appropriate telluric spectrum with the aid of
a custom code; this determined optimal scalings of small regions of
the telluric spectrum to match the corresponding telluric lines in the
science spectrum, and then fitted these with a smoothly-varying   
function to give optimal scalings at every wavelength for the telluric
spectrum.  The telluric-removed science spectra were then normalised
to the continuum.

Where multiple observations had been made of a science target within
the same epoch (in practice, within a few hours of each other on the
same night) they had barycentric corrections applied, and were then
median-combined.  A search for radial velocity variability between
epochs in the brighter cluster members, described fully in paper III,
was then carried out, and where variability was detected, all epochs  
were shifted to a common velocity (the mid-point of the full range of
measured velocities), and then median-combined.  Fainter stars, and
those which were only observed on a single epoch, or for which radial
velocity variability could not be reliably measured, were directly   
median-combined.  All final combined spectra were checked, and
occasional residual bad features manually removed.

SINFONI provides a resolving power (in the $K$-band and at our plate  
scales of 0\farcs1 or 0\farcs025) of $R\sim4500$ at 2.2~$\mu$m.  All
spectra were rebinned to a common dispersion of 0.000245~$\mu$m
pixel\textsuperscript{-1} and common wavelength range of
2.02--2.45~$\mu$m.

\begin{figure*}
\includegraphics[width=13cm,angle=0]{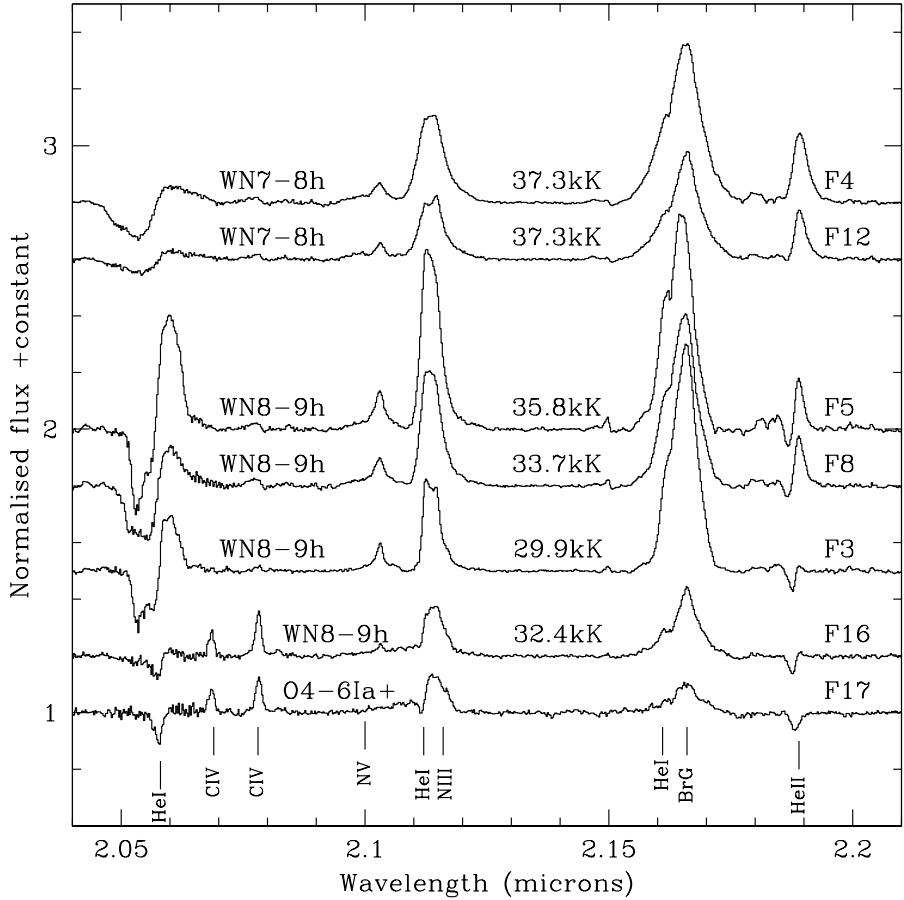}
\caption{K-band spectra of selected WRs ordered on the basis of the
morphology of the He\,{\sc ii} 2.189$\mu$m line. Spectral types
and temperatures inferred by Ma08 indicated.
The spectrum of F17, which appears intermediate between the O4-6 
Ia$^+$ stars and the least extreme WN8-9h star, F16, is shown for 
comparison.}
\end{figure*}

\begin{figure*}
\includegraphics[width=13cm,angle=0]{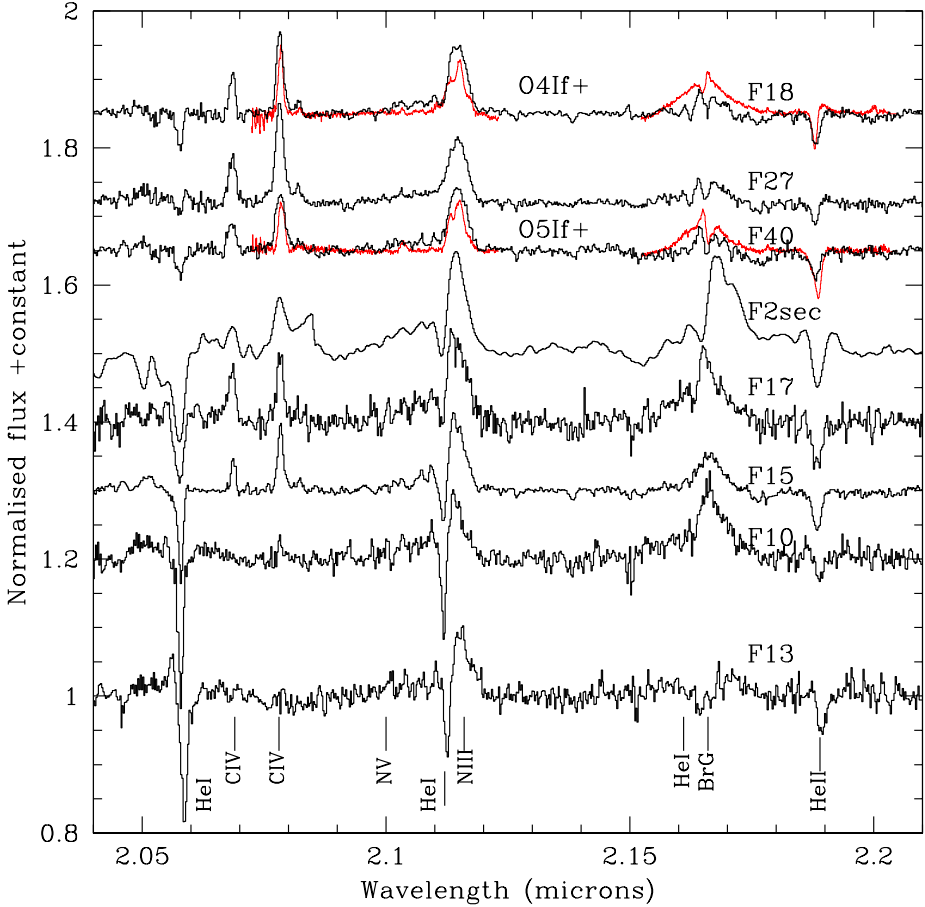}
\caption{K-band spectra of potential extreme O-type  supergiant or hypergiant cluster members (black)
compared to appropriate template spectra from Hanson et al. (\cite{hanson05}; red).}
\end{figure*}

\begin{figure*}
\includegraphics[width=13cm,angle=0]{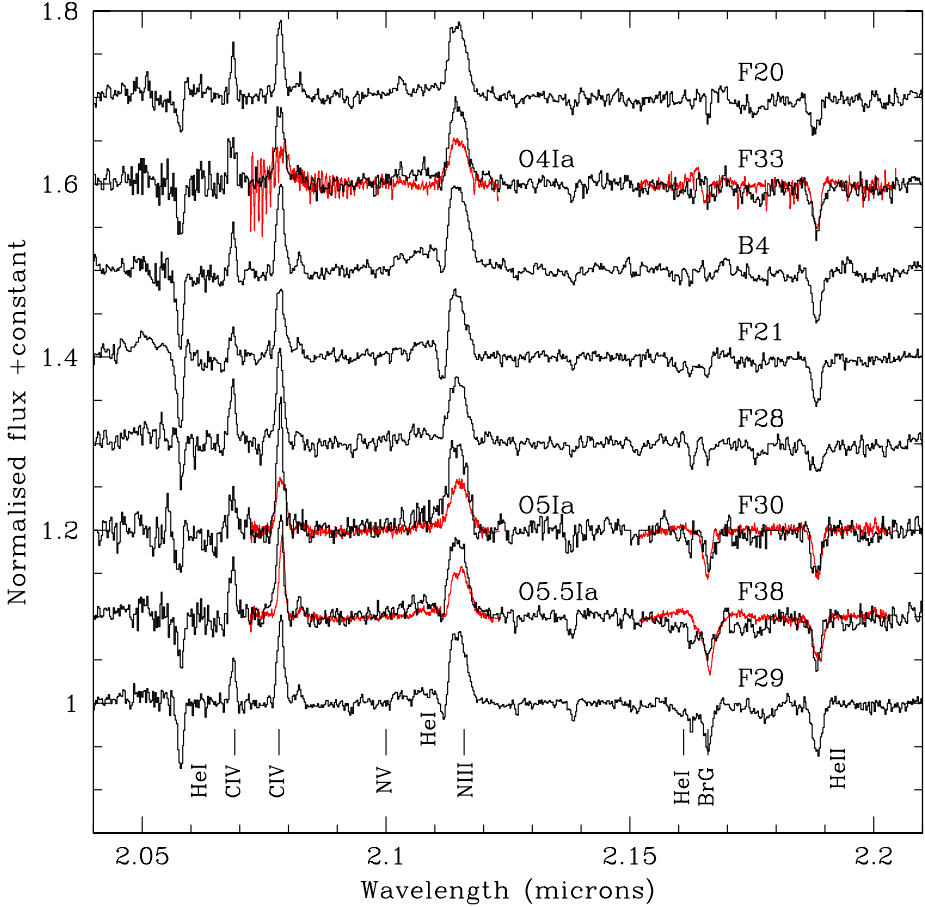}
\caption{K-band spectra of O supergiants within the Arches previously 
sampled by Ma08 (black) compared to template
spectra from Hanson et al. (\cite{hanson05}; red).}
\end{figure*}

\begin{figure*}
\includegraphics[width=13cm,angle=0]{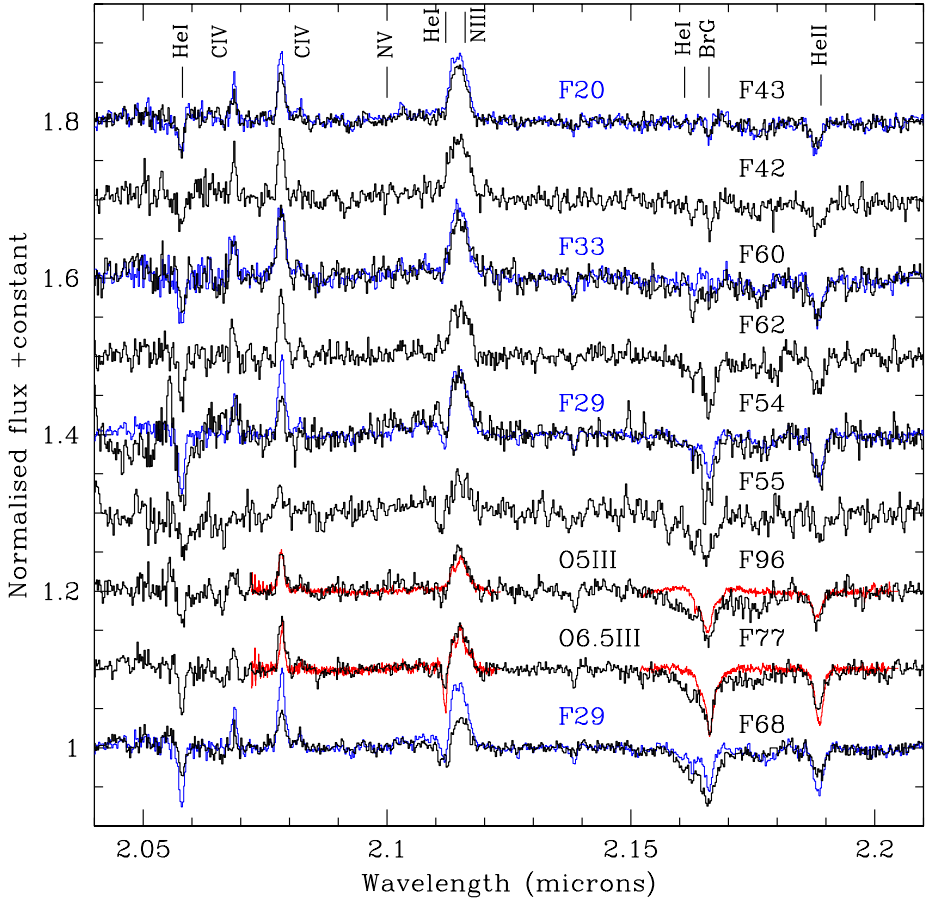}
\caption{Comparison of cluster members (black) to previously classified cluster supergiants (blue)
and  mid O giant template K-band spectra from Hanson et al. (\cite{hanson05}; red). 
Note that no template spectra earlier than O5 III are available.}
\end{figure*}

\begin{figure*}
\includegraphics[width=13cm,angle=0]{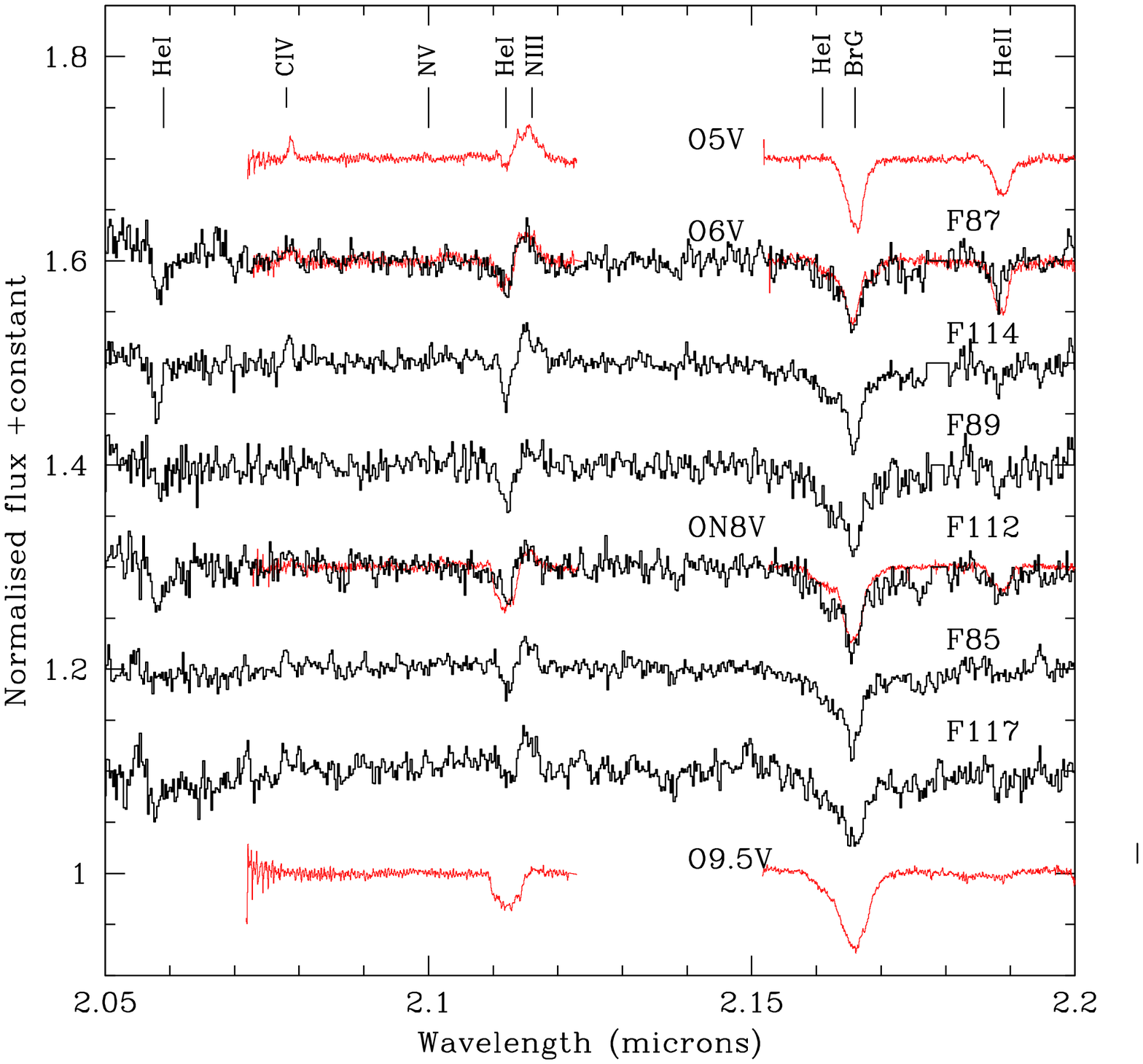}
\caption{Comparison of selected fainter cluster members (black)  with MS K-band template spectra 
from Hanson et al.  (\cite{hanson05}; red).}
\end{figure*}

\begin{figure*}
\includegraphics[width=13cm,angle=0]{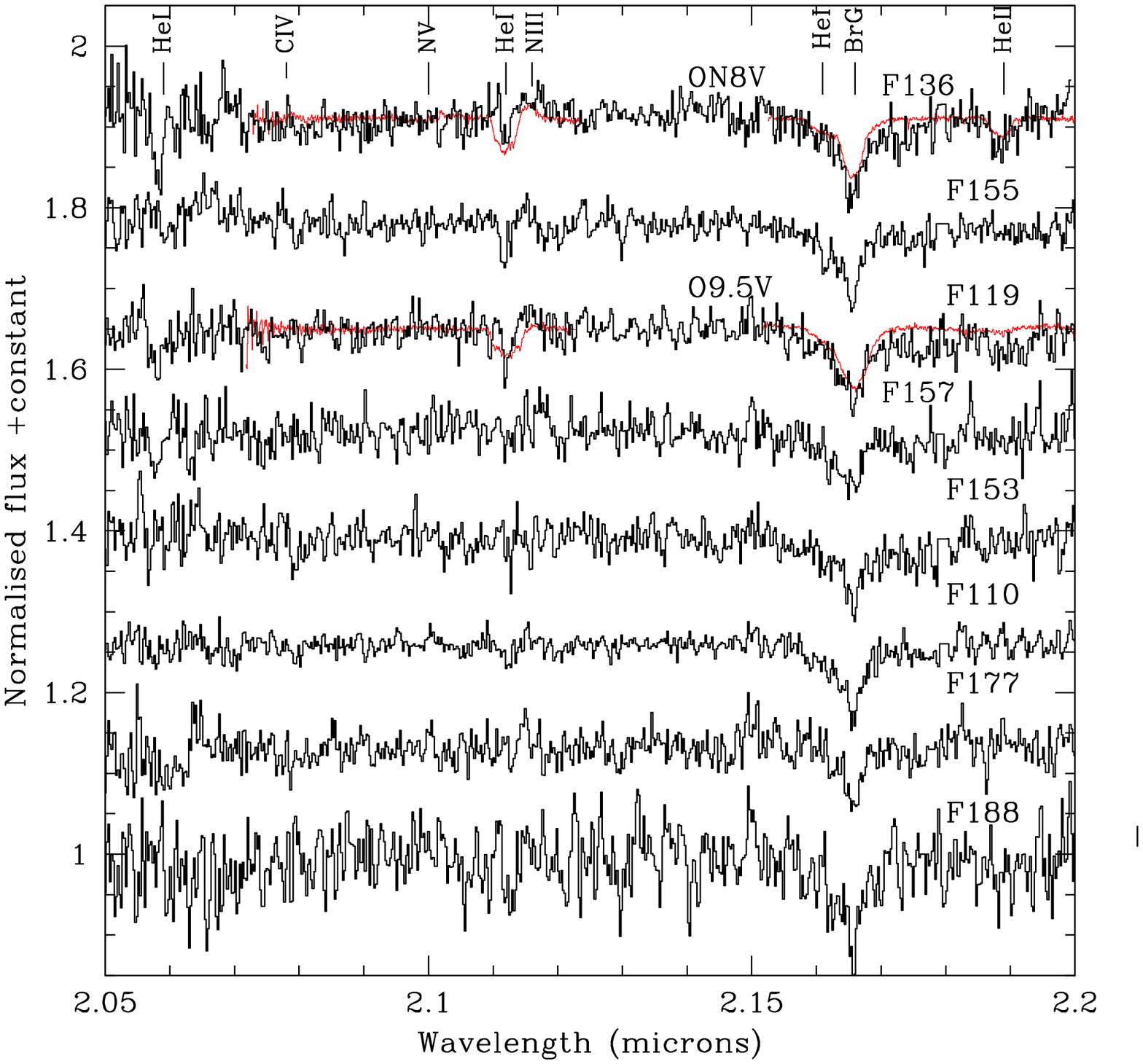}
\caption{Comparison of selected fainter cluster members (black)  with MS K-band template spectra
from Hanson et al. (\cite{hanson05}; red).}
\end{figure*}

In total we extracted spectra for 105 objects. Excluding four apparent late-spectral type interlopers we present the resultant 
spectra of all cluster members  in Fig. A.1, where, following Fi02, they are ordered by decreasing  estimated 
absolute K-band magnitude. Of these 88 appear to be both cluster members and 
of suitable S/N to attempt classification (Sect. 3), more than tripling the spectroscopic sample 
presented by Ma08 and critically extending the census to lower luminosity objects.
Representative spectra illustrating each spectral type and/or  luminosity class are presented in Figs. 1-6.

\subsection{Photometry}
Despite the availability of ground-based AO observations, we have chosen to exclusively employ HST 
data due to its high angular resolution and sensitivity, stability of point spread function and
 accurate zeropoint; invaluable given the compact nature of the Arches.  As well as employing 
photometry from Fi02 (F110W, F160W and F205W) and Dong et al. (\cite{dong11}; NIC3 F190N) we 
present new WFC3 photometry in the F127M, F139M and F153M filters. 
The relevant data were obtained in 2010-12 under programs GO-11671, 12318 and 12667 (PI Andrea 
Ghez). A detailed description of data acquisition, reduction and analysis, including error 
determination, is presented by  Dong et al. (\cite{dong17}), which, for convenience, we reprise here.

Raw data and calibration files were downloaded and the latest HST pipeline, OPUS version and 
CAL-WFC3 version 2.1 were used to perform basic calibration steps on individual dithered 
exposures such as bias correction and flat fielding. Under PyRAF the `Tweakier' and `Astrodrizzle' 
tasks aligned individual exposures and corrected for distortion and masking out defects 
before combining images.

The `DOLPHOT' package\footnote{http://americano.dolphinsim.com/dolphot/} (Dolphin et al. 
\cite{dolphin}) was employed  to extract photometry. DOLPHOT returns photometric uncertainty 
resulting from Poisson fluctuations produced by electrons in the camera and also allowed us to 
 perform additional artificial star tests on the dithered images to provide a secondary, 
parallel error determination. For each star we used the larger resulting error estimate. 
We present error estimates as a function of magnitude for each filter in Fig. 7; with the
 exception of a handful of examples these are less than $\sim$0.05mag in each band. 

The resultant photometry, along with the pseudo K-band F205W magnitudes from Fi02 are 
presented in Table A.1. 

\section{Spectral classification}

 Our observations sample stars with K-band (F205W)  magnitudes ranging from $\sim10.4$ (F6) through to $\sim14.5$ (F155) and 
$\sim15$ (F188). Foreshadowing  the discussion of interstellar extinction in Sect. 4, Fi02 estimate these correspond to   $M_K 
\sim -8$ - $-3.1$. In total we extracted  spectra of 105 individual stars. Of these, spectra of four 
stars\footnote{F11, F46, F51 and F99.} showed pronounced CO bandhead absorption, marking them as M star 
interlopers. The spectra of a further five stars\footnote{F44, F73, F80, F166 and F170.} may potentially be  
contaminated by bright neighbours and hence are not considered further.
 A number of fainter objects\footnote{F159, F168, F173, F174 and  F184.} were 
of sufficiently low S/N that accurate spectral classification was impossible. However these are retained for the purpose 
of discussion since their essentially featureless spectra are {\em not} consistent with identification as e.g. evolved 
Wolf-Rayets, since the broad and strong emission lines expected for such objects would have been readily identifiable. Finally 
three 
further stars demonstrated essentially featureless spectra\footnote{F151 and F176 and F189.} despite having a sufficient S/N 
to identify the classification diagnostics for e.g. main sequence stars if present.
 This  left a total of 88 stars for which spectral classification could be attempted, to which  we may also add the 
reconstructed spectrum of the secondary in the massive binary system F2 (Paper II).

\subsection{Classification criteria}

A number of authors have studied the utility of near-IR spectroscopy for classification of early-type stars; 
specifically Hanson et al. (\cite{hanson96}, \cite{hanson05}) and Crowther \& Furness (\cite{crowther08}) provide quantitative 
criteria for OB stars, with Crowther et al. (\cite{crowther06}) and
Rosslowe \& Crowther (\cite{rosslowe}) replicating these efforts for WRs and Crowther \& 
Walborn (\cite{crowther11}) extending these studies to extremely luminous stars sharing properties of both classes.
Drawing on these studies where available, Ma08 discuss and present  classification criteria for  stars within the 
Arches in some depth; for consistency we adopt their scheme for this  study, revising  their spectral types and/or 
luminosity class assignments only when suggested by the improved S/N of the data presented here. 

However, the increased integration times afforded by stacking multiple spectra of individual objects results in 
qualitatively different spectral morphologies amongst the fainter Arches cohort. Consequently it is worth revisiting 
the classification criteria afforded by the K-band. Prime temperature diagnostics include the He\,{\sc ii} 
2.189$\mu$m line and the absorption line associated with the He\,{\sc i} 2.112$\mu$m component of the broad blend 
resulting from transitions from He\,{\sc i}, N\,{\sc iii} C\,{\sc iii} and O\,{\sc iii}  (cf. Hanson et al. 
\cite{hanson96}, \cite{hanson05}), with the  C\,{\sc iv} transitions between 2.07-2.08$\mu$m providing a secondary 
criterion for stars between O4-8 (peaking at $\sim$O5 and decreasing either side). 

An additional diagnostic for early-type stars that is not discussed in the above works is the strong emission feature
 at $\sim2.42\mu$m that is, subject to sufficient S/N, ubiquitous for the vast majority of our spectra. Non-LTE model atmosphere simulations utilising the CMFGEN code (Hillier \& Miller \cite{hillier98},\cite{hillier99}) suggest the identity of this feature is sensitive to temperature, with contributions from, respectively, the n=$10{\rightarrow}9$ 2.436-8$\mu$m lines  of O\,{\sc iv}, N\,{\sc iv}, C\,{\sc iv} and finally Si\,{\sc iv} as one transitions to cooler temperatures.
Since the WNLh stars considered here are likely $>30$kK we would expect Si\,{\sc iv} to dominate the emission feature. N\,{\sc iv} starts to contribute at $\sim32$kK, equalling the strength of Si\,{\sc iv} at $36.5$kK.
 Subject to depletion C\,{\sc iv} might be expected to play a minor contribution; since these stars are almost certainly cooler than  $\sim45$kK one would not expect a contribution from the O\,{\sc iv}. Similarly for supergiants of spectral type mid-O and later  one would expect this feature to result from a combination of Si\,{\sc iv}, C\,{\sc iv} and N\,{\sc iv}, with Si\,{\sc iv} increasingly dominant for cooler stars.

Br$\gamma$ is another key diagnostic line, providing valuable information on both the mass-loss rate and  the luminosity 
class of the star via the observation that, in general, increased luminosity is associated with elevated mass-loss rates. Such a 
relationship appears realised in the Arches, where the line is seen fully in emission in both the WR and hypergiant cohorts. Less-evolved stars 
support lower stellar luminosities and are  normally associated  with reduced mass-loss rates, favouring a transition to photospheric absorption profiles and  motivating the division 
between hypergiants and supergiants adopted by Ma08. Once seen in absorption, the shape of the photospheric profile may in 
theory be used to assign a luminosity class via the  dependence on surface gravity (Hanson et al. \cite{hanson05}).
However, given the luminosities of the stars in question in this study,  this may be  complicated by residual  
contamination of the profile by wind emission and rotational broadening (although the latter is typically dominated by the large intrinsic widths of the photospheric absorption wings).

Once the additional temperature dependence  of the H\,{\sc i} and He\,{\sc i} transitions is taken into account it becomes 
difficult to finely discriminate between adjacent luminosity classes (e.g. Ia, Ib and II and III-V). This is illustrated in Fig. 
8, where we compare template spectra for giant and main sequence objects from Hanson et al. (\cite{hanson05}). For mid-O (O5-6) 
spectral types the strength of C\,{\sc iv} emission distinguishes between  luminosity types, being sytematically stronger in the 
giants. However  despite the excellent S/N ratio ($>100$) of the spectra, once this diagnostic disappears at lower temperatures it 
becomes more problematic to reliably differentiate between $\sim$O7-O9 giant and main sequence stars, since in isolation one is 
forced to  rely solely on the wings of the Br$\gamma$ photospheric line. 

Nevertheless,  for a subset of high S/N spectra - exemplified by  F68, F77 and  F96 (Fig. 
4) - a classification as giant appears most appropriate given both the marked similarity to  the spectroscopic standards  of 
Hanson et  al. (\cite{hanson05}) and notable  differences compared to the `{\em bona fide}' Arches supergiants (cf. discussion in 
Sect. 3.2). Likewise we may identify a number of candidate early-mid O dwarfs amongst the photometrically faintest cluster members 
- e.g. F87 and F112 (Fig. 5) - that  differ  from the assumed giant cohort due to the comparative weakness of C\,{\sc iv} 
emission. Both 
sets of spectra may hence be employed as `anchor points' from which one may  `bootstrap' classify other similar but lower S/N 
spectra and, in the  case of the main sequence candidates, stars of later spectral type, which demonstrate a smooth morphological 
progression. A similar  approach was also adopted for the more luminous super- and hypergiants within the Arches.
We discuss this process in more detail below, cautioning that this still 
leaves a number of stars for which an  indeterminate - e.g. I-III or III-V - luminosity class is most appropriate given the 
quality of the current dataset.

\subsection{Analysis of the dataset}

The principal conclusion from consideration of the whole dataset is that there is a remarkable similarity and 
continuity between the spectra of individual stars within the Arches as one progresses from higher to lower 
luminosity objects. While we are able to identify qualitatively distinct spectral morphologies amongst the fainter cluster 
members  when compared to the more luminous subset presented by Ma08, the  population of the Arches that we sample appears to 
consist predominantly, and possibly exclusively  of WNLha and O  stars.
As described below, despite reaching a magnitude at  which they should be detectable, we are unable to detect  any more evolved 
Wolf-Rayets in the cluster, such as H-depleted WNE or WC stars. Likewise  transitional objects such as luminous blue variables 
(LBV) and  blue hypergiants of later spectral type are also absent. These findings have important  consequences for the age of the 
Arches, which we return to in Sect. 5.

We emphasise at this point that the spectral and luminosity classifications discussed below are based {\em solely} on spectroscopic 
data. Unfortunately, the  absolute magnitudes of individual stars, which would aid classification, are dependent on the highly  
uncertain extinction law towards the Arches and the differential reddening evident across the cluster. Indeed it is hoped that 
identification of stars with a well defined luminosity, such as the main sequence, will help refine these parameters: an  issue we 
return to in Sect. 4.

\subsubsection{WRs and O-hypergiants}

With Arches F11 identified as a foreground  interloper, we fail to identify any further WRs within the Arches, while the 
higher S/N spectra provide no compelling evidence to revisit their spectral classifications. A subset of spectra are plotted in 
Fig. 1 and  following Crowther et al. (\cite{crowther06}) and
Crowther \& Walborn (\cite{crowther10}) are  ordered
primarily by the strength and morphology of the He\,{\sc ii} 2.189$\mu$m
line, the ratio of which to Br$\gamma$ serves as a classification
criterion for WNLh stars. As anticipated this broadly correlates
 with a decrease in the  temperatures for individual stars found by
Ma08. Considerable diversity is present in the spectral morphologies
of these stars, most notably  in the shape of the He\,{\sc i} profile and the 
 strengths of the nitrogen, carbon and oxygen transitions, which Ma08
attribute  to differences in CNO burning
products at the stellar surface.

Ma08 identify two extreme O super-/hypergiants - F10 and F15 -  within the Arches on the basis of strong Br$\gamma$ emission,  a 
criterion also consistent with examples presented in the spectral atlas of Hanson  et al. (\cite{hanson05}).  Following  this 
criterion we significantly expand  on the number of O hypergiants present in the cluster, identifying a further six 
candidates - 
the secondary in the binary F2, F13, F17, F18, F27 and  F40 (Fig. 2).

We highlight the morphological  similarities between the most extreme example, F17, and the least extreme 
WNLh within the cluster, F16, which suggest a close evolutionary connection between the two classes of star (Fig. 1). Indeed the 
increased 
population of hypergiants within the Arches strengthens the hypothesis by Ma08 that they represent an intermediate evolutionary phase
between the supergiant and WR populations.

Of the hypergiants, comparison of F18, F27 and F40 to the template spectra of Hanson et al. (\cite{hanson05}) suggest that they 
are O4-5 
If$^+$ stars, albeit with slightly lower mass-loss rates inferred from the weaker Br$\gamma$ emission. Conversely the secondary in the 
eclipsing binary F2 along with F10, F15 and F17 demonstrate much stronger Br$\gamma$ emission along with the presence of a notable 
absorption component to the 2.11$\mu$m blend attributable to the He\,{\sc i} 2.112$\mu$m  transition (behaviour mirrored in the 
He\,{\sc i} 2.059$\mu$m singlet).
In conjunction with the  reduction in strength of the C\,{\sc iv} 2.069$\mu$m and 2.079$\mu$m lines -  the object with the deepest 
He\,{\sc i} 2.112$\mu$m  feature, F10, also shows the weakest C\,{\sc iv} emission - this observation suggests these are cooler 
objects than  F18, F27 and F40, albeit with  denser winds.  Given the lack of 
suitable spectral templates we {\em provisionally} classify these as  O5-6 Ia$^+$ (F2 secondary and F17), O6-7 Ia$^+$ (F15) and O7-8 
Ia$^+$  (F10), noting that quantitative model-atmospheric analysis will be required to confirm this spread in temperatures.

Finally we turn to F13 which, despite sporting weak  Br$\gamma$  absorption demonstrates 
 deep absorption in the He\,{\sc i} lines  and a lack of C\,{\sc iv} emission -  a combination of features not replicated in {\em any} 
other cluster member with the exception of  F10. Given this, and mindful of the Br$\gamma$ absorption we therefore adopt a similar 
classification
of  O7-8 Ia$^+$.

\subsubsection{O supergiants}
Next we consider the O supergiants within the Arches, concentrating first on the brighter cohort considered by Ma08,  
comprising 13 stars between F18-F40, to which we add a further six stars\footnote{B4, F19, F24, F25, F30 and 
F38.}, all of which appear to be O supergiants. Attribution of absolute spectral types is complicated by the lack of suitable 
template  spectra, 
particularly  for stars of  the  luminosities expected for the Arches. In this respect  we highlight that comparison to the 
spectra of Hanson et al. (\cite{hanson05}) reveals that emission lines in cluster supergiants are systematically 
stronger (cf. F30 and F33; Fig. 3). Nevertheless the higher S/N afforded by the new data allows us to provide robust relative 
calibrations since we can now reliably identify systematic variations in both 
Br$\gamma$ (a proxy for mass-loss rate/wind density) and He\,{\sc i} 2.112$\mu$m (temperature) lines (Fig 3).

While the majority of these stars show the 2.11$\mu$m blend purely in emission, indicative of an O4-5 Ia classification 
(e.g. F20, F28 and F33; Figs. 3 and A.1 and Table A.1), five stars show He\,{\sc i} 2.112$\mu$m absorption components of 
various strengths, suggestive of later spectral types; we assign provisional classifications of  O5.5-6 Ia (B4, F22 and F29) and 
O6-6.5 Ia (F21 and F23). Irrespective of spectral type we identify varying degrees of infilling of the He\,{\sc i}+Br$\gamma$ 
photospheric blend, which is  almost absent in  B4, F20, F21 and  F33 (Figs. 3 and A.1). This would suggest a close connection 
with the hypergiants -  where this blend is seen in emission - for these stars, with F20 and F33 (no He\,{\sc i} 2.112$\mu$m 
absorption) being direct 
antecedents of hotter O hypergiants such as F18, F27 and F40, while  B4 and F21 (He\,{\sc i} 2.112$\mu$m 
absorption) fulfil
this role for the cooler hypergiants F10, F15 and F17. 

We now turn to fainter stars for which no previous spectra have been published; these would be expected to 
comprise lower luminosity supergiants, giants and possibly main  sequence objects. We present a montage of spectra of selected  
stars in Fig. 4 and it is  immediately clear from comparison to known supergiants that additional examples are present within 
this cohort - e.g. F42, F43 and F60. In total we identify a further 8 supergiants, noting that none have spectral types outside 
the previous range  (O4-6.5 Ia; Table A.1).  We caution that following Fi02, the increasing `F' numbers of these stars is 
indicative of reduced  luminosities and so  luminosity classes Iab, b or II may in reality be more appropriate for some than the 
generic Ia assigned here. Indeed this possibility is reflected in the assignment of a luminosity class I-III for six stars for 
which emission lines are significantly weaker than {\em bona fide} supergiants (cf. F54 and F62 in Fig. 4, Table A.1). 

To summarise: Ma08 identified 13 mid-O supergiants of which we reclassify two as hypergiants. We  more than double this number,
finding a total of 25 O4-6.5 Ia supergiants within the Arches, with a futher six objects of comparable spectral type  assigned 
a luminosity class I-III to reflect a systematic reduction in emission line strength accompanied by an increase in strength of 
the Br$\gamma$ photospheric line.

\subsubsection{O giants and main sequence stars}

As discussed in Sect. 3.1 (and illustrated in Fig. 8), given the difficulty in distinguishing between main sequence and giant O 
stars we consider the remaining objects together. As with more luminous
objects we employ the presence or otherwise of He\,{\sc i} 2.112$\mu$m and He\,{\sc ii} 2.189$\mu$m absorption in conjunction with
C\,{\sc iv} 2.069+2.079$\mu$m  emission (where present) to constrain spectral type, while 
 luminosity class diagnostics are outlined in Sect. 3.1.
Direct comparison of observations to template spectra identifies six potential
 O giants (Figs. 4 and A.1) with spectral types ranging from $\sim$O5-6 (F96) to $\sim$O6-6.5 
(F77). This conclusion is  bolstered by comparison of the spectrum of  F68 (O5.5-6 III) to the  supergiant  
F29 (O5.5-6 Ia; Fig. 4); emission 
lines are systematically weaker while the Br$\gamma$ photospheric profile is deeper, with much more pronounced wings. 
 We assign an intermediate   (III-V) luminosity class to a further five stars (spanning the same range of spectral types) where 
the comparatively  low S/N compromises an 
assessment of the strength of the intrinsically weak C\,{\sc iv} 2.079$\mu$m line  relative  to the  2.11$\mu$m emission blend.

We also identify a  population of stars which we tentatively classify  as main sequence stars on the basis of the 
width and depth of the Br$\gamma$ line and, for the earliest spectral types, the absence of C\,{\sc iv} emission even when other
 indicators of high temperature - such as strong He\,{\sc ii} 2.189$\mu$m absorption and a lack of a He\,{\sc i} 2.112$\mu$m 
absorption component in the $\sim2.11 \mu$m emission blend  - are present (cf. Fig. 8). Examples are presented 
 in Figs. 5 and 6, with the earliest examples being O5-6 V (e.g. F87). However the relatively low S/N 
of many of these 
spectra  complicates  identification of features such as  He\,{\sc ii} 2.189$\mu$m  and so greater 
uncertainty is associated with these spectral 
classifications.  Indeed in a number of spectra Br$\gamma$ is the only line which may be confidently identified and hence we adopt 
a generic $\geq$O8 V classification for such stars, with the expectation that a  number of stars may be 
substantially later (e.g. 
early-B) than this.

For completeness we highlight anomalies in the Br$\gamma$ profiles of a number of stars (Fig. A.1). F81 (O6-7 
III-V) and,  
subject to the low S/N of the spectrum, F139 appear to show 
a sharp central emission peak superimposed on a broad photospheric profile. Likewise infilling leads to an essentially flat 
continuum for  F90 and  F92 (both O5-6 V) and F93 (O5-6 III-V). Further observations will be required to assess the veracity of 
these features.

\subsubsection{The absence of H-free WRs, LBVs and interacting binaries}
 
Finally, we consider the lack of certain types of evolved objects within the Arches. Turning first to H-free WRs, and if
 we are correct in  our detection of the main sequence within the Arches, then we are reaching stars with 
$M_{K}\sim-4.4$(O5 V) to  $\sim-3.3$ (O9 V; Martins \& Plez \cite{martins06}). Consequently,
the empirical values of  $M_K$ presented for both WN and WC field stars in Crowther et al.
(\cite{crowther06}) imply  that, with the exception of the weak lined WN3-4 stars, we would
expect to detect any H-free WN or WC stars within the Arches. Similar conclusions may be
obtained upon  consideration of the absolute near-IR magnitudes of the WR cohort within Wd1
(Crowther et al.  \cite{crowther06}). 

Theoretical simulations support such an expectation. Groh et al. (\cite{groh13}) present $M_K$s for the pre-SN (WC or
WO) endpoints of single massive (60-120$M_{\odot}$) stars which are in the range ($M_{K} \sim
-3.3 - -4.5$) that we sample. Secondly Groh et al. (\cite{groh14}) present a detailed appraisal 
to the lifecycle of a single  $60M_{\odot}$ star and show that after spending $\sim3.2$ Myr as 
an O supergiant it evolves through a variety of high luminosity phases (blue hypergiant and  
luminous blue variable) before reaching H-free WN, WC and finally WO states - all of which
would be readily detectable in the Arches and are significantly longer-lived than the  WNLh phase.

However even if we are mistaken in our identification of the cluster main sequence, comparison of the Arches to 
the Quintuplet cluster is strongly suggestive of a lack of H-free WC stars. Located at a comparable 
distance and presumably observed through a similar extinction column, the Quintuplet has a rich population of WC stars 
(e.g. Liermann et al. \cite{liermann09}). Of these, the faintest, potentially single stars have $m_{\rm F205W} \sim
11.3 - 11.7$mag, while those found in binaries are signficantly brighter still, ranging up to $m_{\rm F205W} \sim 7.2$mag (due to
 the formation  of hot dust in the wind collision zone; Clark et al. submitted). In comparison we have spectral classifications for all stars to a limit of $m_{\rm F205W} \sim 12.41$mag (F31, Table A.1), strongly suggesting that no WC stars are present within the Arches at this time unless subject to particularly extreme differential reddening. Moreover we may  not easily appeal to the presence of such  a  star 
in a binary in which emission from the  companion overwhelms it unless it is in such an orbital   configuration that dust does 
not form {\em and} it is intrinsically  fainter than examples within the Quintuplet.  Trivially, similar conclusions may also 
be drawn from consideration of the massive field star population of the Galactic Centre (Mauerhan et  al. \cite{mauerhan10a}, \cite{mauerhan10b}) and the Galactic Centre cluster (Martins et al. \cite{martins07}).

Given the intrinsic IR luminosity of (candidate) luminous blue variables (LBVs) and cool hypergiants we can be sure that
none are present within the Arches. Regarding LBVs; while it might be supposed that the Arches is  too young for this 
phase to be encountered, the (pathological) LBV 
$\eta$  Carina is located within the Trumpler 16/Collinder 228 stellar aggregate, which Smith (\cite{smith06}) show 
to contain a comparable stellar population of H-rich WNL and O stars to the Arches. 

Lastly, while X-ray observations and our RV studies suggest that a number of (colliding wind) binaries are found  within the cluster (Wang et al. \cite{wang}, Papers II and III), the Arches appears to lack any systems in which rapid case-A mass-transfer is ongoing. 
Several examples of such systems have been proposed - Wd1-9 (Clark et al. \cite{clark13}, Fenech et al. \cite{fenech}), RY Scuti (Gehrz et al. \cite{gehrz95}, \cite{gehrz01}, Grunstrom et al. \cite{grundstrom}), NaSt1 (Mauerhan et al. \cite{mauerhan15}) and  LHA 115-S 18 (Clark et al. \cite{clark13a}) - and appear to manifest as (supergiant) B[e] stars (cf. discussion in Kastner et al. \cite{kastner}), supporting a combination of a rich low excitation emission line spectrum and  bright X-ray, IR and sub-mm/radio continuum emission (due to colliding winds, the presence of hot dust and a highly elevated mass-loss rate, respectively). Such a combination of observational features would render such stars readily identifiable within the Arches, but none appear present.

\begin{figure}
\includegraphics[width=7.5cm,angle=0]{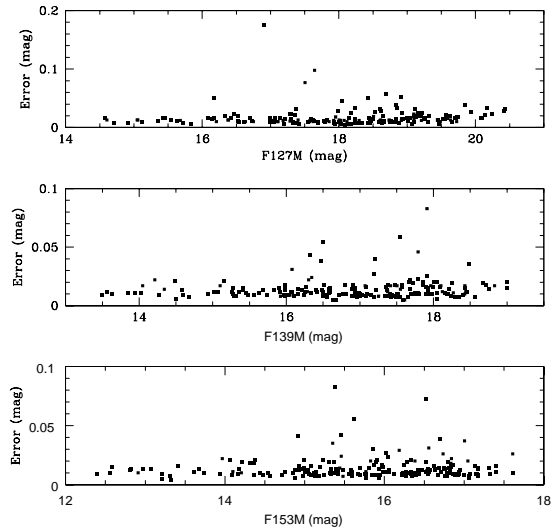}
\caption{Errors associated with the HST/WFC3 F127M, F139M and F153M photometry presented in Table A.1.}
\end{figure}

\begin{figure}
\includegraphics[width=7.5cm,angle=0]{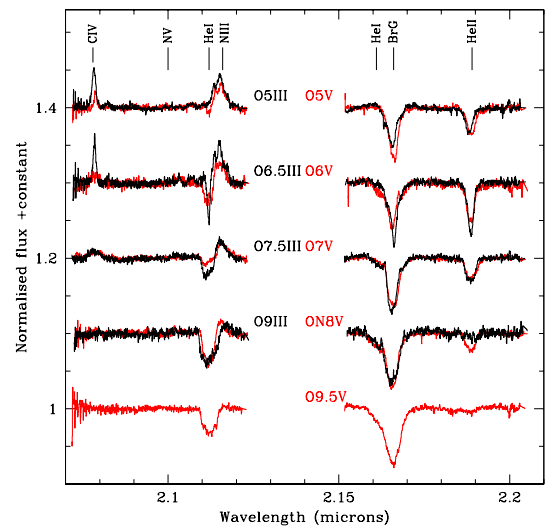}
\caption{Comparison of K-band classification spectra from Hanson et al.
(\cite{hanson05}) for mid-late O giant (black) and main sequence
(red) stars. For earlier spectral types the strength of C\,{\sc iv} relative to e.g. the 2.11$\mu$m emission
blend serves to distinguish between giants and main sequence stars, being weaker in the latter. For later
spectral types one must rely on the shape of the Br$\gamma$ profile, although this can be compromised by
 stellar   rotational velocity.}
\end{figure}

\section{The impact of an uncertain extinction law towards the Galactic centre}

During the classification of cluster members we have intentionally excluded consideration of magnitudes 
due to the uncertainty in the correct interstellar reddening law to apply. Indeed we wish to utilise the main sequence 
cohort 
to better constrain reddening and so utilising photometric data in their identification would introduce a circularity into the 
argument. 

Three issues beset attribution of interstellar reddening to individual stars: adoption of the correct reddening law, the presence 
of inhomogenous reddening across the cluster and an intrinsic IR-excess due to  continuum emission from  stellar winds. A number of 
studies have attempted to determine the reddening law to the Galactic centre, with 
Rieke \& Lebofsky (\cite{rieke}) providing the first optical-IR constraints. Their results were then  generalised to a power-law 
formulation ($A_{\lambda} \propto \lambda^{-{\alpha}}$  with  $\alpha=1.61$) by Cardelli (\cite{cardelli}), although later 
studies 
utilising near-IR surveys (Nishiyama et  al. \cite{nishiyama}) and red-clump stars (Schoedel et al. \cite{schoedel}) suggest a 
steeper index of  $\alpha=2$.  An alternative, derived for the nearby Quintuplet cluster from analysis of mid-IR data and based 
on the work  of Lutz et al. (\cite{lutz}),  was described by Moneti et al. (\cite{moneti}); given the proximity of both clusters 
such a 
prescription could be appropriate for the Arches too. 

Habibi et al. (\cite{habibi}) studied the reddening towards the Arches based upon the power-law formulation of Rieke 
\& Lebofsky (\cite{rieke}; $\alpha=1.61$) and Nishiyama et al.  (\cite{nishiyama}; $\alpha=2$)  and demonstrated  that 
extinction across the cluster is highly variable with no strong systematic trends as a 
function of location, ranging from 
$2.7<A_K<4.5$mag and 
$2<A_K<3.4$mag respectively. This has the potential to lead to large uncertainties in luminosity  even 
before uncertainties in the intrinsic colour of individual  stars due to wind emission are considered.

To emphasise the difficulty in constraining interstellar reddening and hence stellar luminosities, in  Figs. 9 and A.2 we present the results of preliminary analysis of four  of the WNLh stars, which are known to support strong stellar winds. We employ the CMFGEN non-LTE model atmosphere 
code of Hillier \& Miller (\cite{hillier98},\cite{hillier99}) following the methodology described in Najarro et al. 
(\cite{najarro}). Spectroscopic data from the current paper  were utilised as well as photometry from Table 1, Fi02 and Dong et al.  (\cite{dong11}), paying particular attention to the bandpasses of individual filters. We adopted two prescriptions for the reddening law - a power-law for which the index was allowed to vary and that of Moneti et al. (\cite{moneti}). 

The differences between best-fit models employing the two  reddening laws are stark; in each case the power-law prescription results in a luminosity a factor of 3.4 (F7) to 4.5 (F6) times smaller than that derived under the assumption of the Moneti et 
al. (\cite{moneti}) law. While the luminosities derived from the latter are more in line with those derived from Ma08, given this 
gulf we refrain from discussing detailed model results for the physical properties of individual stars. However, more generally,  
we note 
that in each case the  indices for the power-law prescription  differ from one another, which is not expected on a physical basis, and  are 
systematically  steeper than previous derivations, leading to unexpectedly low stellar luminosities. We take both observations as 
 a hint that this formulation does not provide a true description of either the reddening law or the stars 
themselves. However  firm  conclusions must await individually tailored quantitative analyses of a larger, 
statistically robust sample of cluster members.

Moreover such an analysis also emphasises the role of differential reddening across the Arches. For the examples in this work the effect is  limited, with extinction adopting the Moneti (power-law) prescription ranging from $A_K \sim3.43 - 3.62$ for the WNLh stars considered, but Lohr et al. (submitted) report $A_K \sim4.52$ via an identical methodology for the similar object F2 - a range of $\sim$1.2mag. which reflects the results of Habibi et al. (\cite{habibi}).

\begin{figure*}
\includegraphics[width=16cm,angle=0]{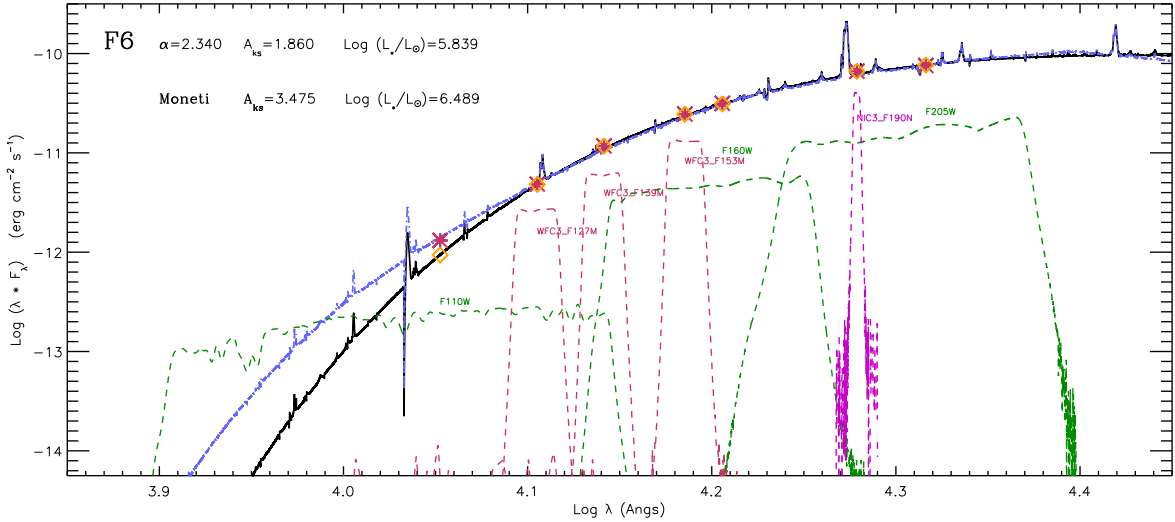}
\caption{Synthetic model-atmosphere spectra for the WNLh star F6 computed for two differing assumed
interstellar reddening laws, illustrating the dramatic dependence of bolometric luminosity on this choice.
HST photometry employed are from Table 1, Fi02 and Dong et al. (\cite{dong11}). The black line reflects the model spectrum 
that was reddened with $\alpha$=2.34 which results in $A_{ks}$=1.860, while the blue line follows Moneti's law with 
$A_{ks}$ =3.475. Transmission curves for the filters 
 used for the fit are shown in green (broadband) and pink (narrowband), and symbols are plotted for each magnitude 
measurement to show the goodness of fit: orange diamonds for the $\alpha$-model and pink stars for the Moneti model.
  The x-axis position of each symbol corresponds to the classical $\lambda_0$ of the filter at which the zero-point 
 flux is defined. The y-axis position coincides with its corresponding model curve if the observed magnitude matches 
 the magnitude of the reddened model. Although the models differ considerably for Log$\lambda<4.1$ (e.g. the F110W 
 filter), they yield the same observed magnitudes. The reason for this is that, due to the high extinction, the 
 reddened-SEDs fall off very steeply as lambda decreases and as a result the effective wavelength of the filter moves 
to the red.}
\end{figure*}

\begin{figure}
\includegraphics[width=7.5cm,angle=0]{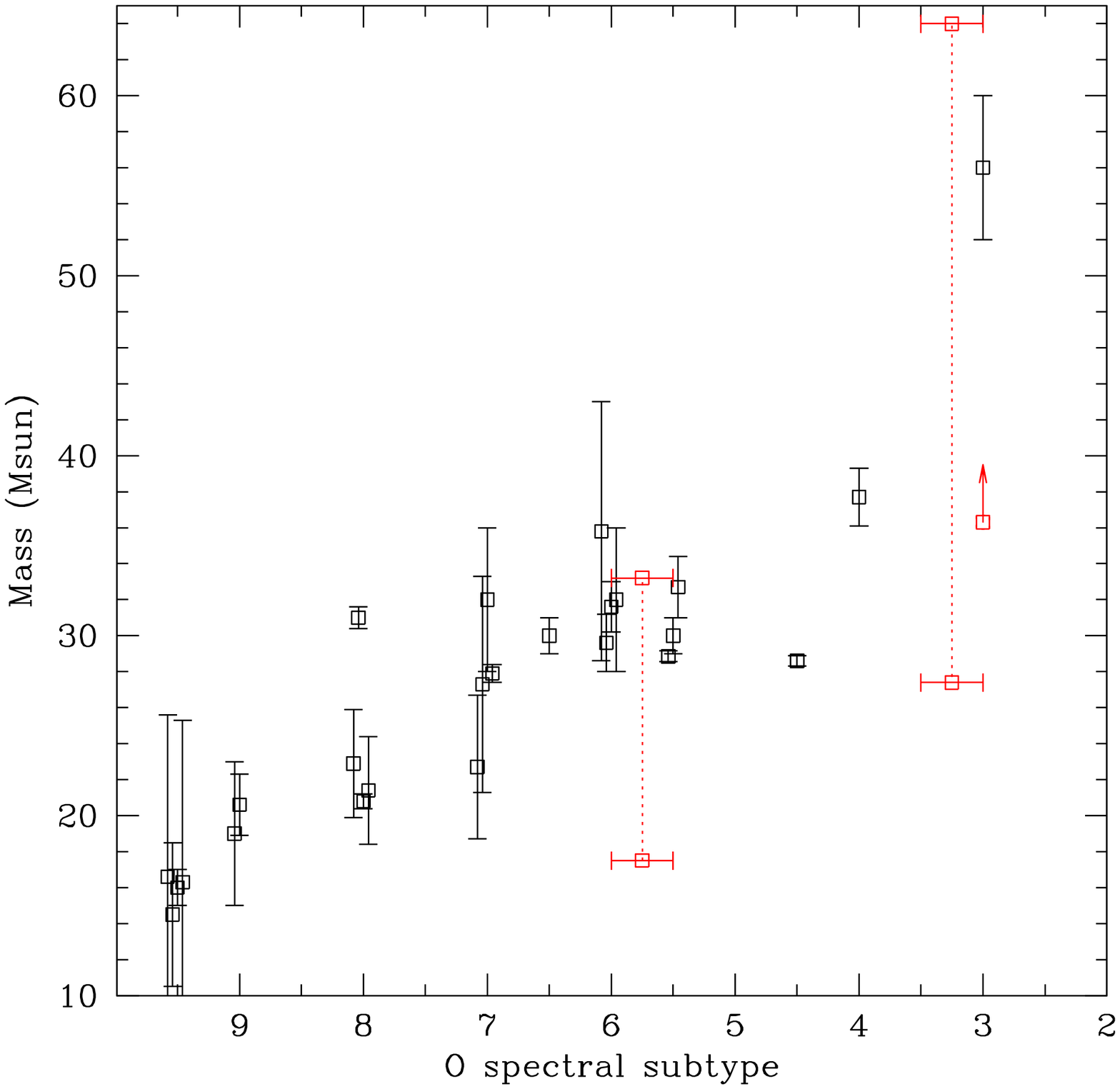}
\caption{Dynamical masses as a function of spectral type for main-sequence O 
stars. Symbols in black denote eclipsing systems as described in Sect. 5.1.1. Symbols in red denote the lower dynamical 
limit for the non-eclipsing secondary in WR21a and the primary and secondary  components of HD 150136, with the dotted lines connecting to spectroscopic
mass estimates for the latter two stars. Error bars for these two objects reflect uncertainties in their spectral type. Small offsets have been applied 
to individual stars of spectral types O5.5, O6, O7, O8, O9 and O9.5 for clarity.
 }
\end{figure}

\section{Stellar and cluster properties}

 Given the significant uncertainties ($\Delta L_{\rm bol} \sim0.6$dex) introduced by the lack of an accepted  extinction law for the galactic centre, the effect of differential  reddening ($\Delta L_{\rm bol} \sim0.4$dex) as well as 
the effects of unrecognised  binarity ($\Delta L_{\rm bol} \sim0.3$dex) we  refrain from constructing a cluster HR diagram at this time. As a consequence  we  are also unable to provide a quantitative age  estimate for the Arches from isochrone fitting, an estimate  of  masses for individual objects via comparison of current physical properties to evolutionary tracks and
an  accurately calibrated cluster luminosity function, or  products derived from it, such as the (I)MF. In particular the errors in luminosity ($\Delta L_{\rm bol} \sim0.2$dex) adopted by both Ma08 and   Schneider et al. (\cite{schneider14}) in their construction of an HR diagram and determination of the cluster (I)MF (respectively) appear to underestimated; leading to greater uncertainty in physical properties derived from these products.

 The lack of such parameters are a serious impediment to exploiting the potential the Arches offers to probe the
lifecycle of very massive stars;  is there a  high-mass truncation to its mass  
function and if so is it imposed by the physics of star formation/evolution or instead  due
to the most massive stars already having been lost to supernovae (SNe)?
 Given the magnitude of the uncertainties associated with current luminosity determinations, it is not 
 immediately obvious how robust any extant age estimate for the Arches is (cf. Sect. 1). Such a situation is 
 regrettable, since  the ages determined by such studies  currently straddle the threshold at which very massive  
 stars may be expected to undergo SNe. 

Despite  these observational limitations, we may still utilise a combination of (i) dynamical mass estimates  via 
cluster binaries and  (ii) observational and/or theoretical 
calibration of   the mapping  between stellar mass and  spectral type/luminosity class to estimate stellar 
masses.  Similarly, age constraints may be imposed by comparison of the  stellar content of the Arches 
 to  (i) those of other clusters for  which age determinations via isochrone fitting have proved possible and 
(ii) the results of  theoretical studies undertaken  to determine the time  at which stars of a  given mass evolve 
through particular evolutionary phases.
Before discussing this methodology in detail we highlight that the theoretical studies employ predictions based on 
 the evolutionary pathways of single stars, which might be significantly modified if instead a binary channel dominates
(cf. Schneider et al. \cite{schneider14}).

\subsection{Masses of cluster members}

\subsubsection{Dynamical masses}

While a number of binary candidates have been identified within the Arches (Paper III), due to the 
limited temporal coverage of our dataset dynamical mass estimates have only been obtained for the 
components of the eclipsing  SB2 system F2 (paper II).  This comprises a WN8-9h primary and O5-6 
Ia$^+$ secondary, with {\em  current} masses of $\sim 82{\pm}12M_{\odot}$ and  $\sim 60{\pm}8M_{\odot}$ 
respectively. Such an extreme mass for the primary is consistent with dynamical estimates for other  
hydrogen-rich WNLh  stars\footnote{e.g. the Galactic stars WR20a  (WN6ha + WN6ha, $83{\pm}5 + 
82{\pm}5M_{\odot}$; Bonanos et al.  \cite{bonanos}), WR21a (O3/WN5ha + O3 Vz((f$^*$)), $>64.4{\pm}4.8 
+ >36.3{\pm}1.7M_{\odot}$; Tramper et al. \cite{tramper}) and NGC3603-A1  (WN6ha + WN6ha, $116{\pm}31 + 
89{\pm}16M_{\odot}$; Schnurr et  al. \cite{schnurr}).}, albeit subject to the twin caveats that these 
examples are  of early spectral subtypes (possibly a function of differing metallicities) and that some lower 
mass examples are also known\footnote{e.g.
WR22 (WN7ha + O8-9.5 III-V,  $55.3{\pm}7.3 + 20.6{\pm}1.7M_{\odot}$; Schweichardt et al. \cite{sch}).}.

There are a similar handful of extreme  O super-/hypergiants with dynamical mass 
determinations\footnote{The Galactic systems  LS III+46 11 (O3.5 If$^*$ + O3 If$^*$, $>38.8{\pm}0.8 + 
>35.6{\pm}0.8M_{\odot}$; Ma{\'i}z-  Apell{\'a}niz et al. \cite{MA}),  Cyg OB2 B17 (O7 Iaf$^+$  + O9 
Iaf, $60{\pm}5 + 45{\pm}4M_{\odot}$; Stroud et  al. \cite{stroud}), the primary of the X-ray binary HD 
153919 (O6.5Iaf$^+$, $58{\pm}11M_{\odot}$; Clark et al. \cite{clark02}) and 
the LMC star R139 (O6.5Iafc + O6 Iaf, $>78\pm8 + >66{\pm}7M_{\odot}$; Taylor et al. \cite{taylor}).}. 
While caution has to  be applied when considering HD 153919 - which is a product of binary 
evolution (Clark et al. \cite{clark02}) - and  the lower metallicity system  R139, the mass of the secondary in 
F2 is  also fully consistent with these values; as with the WNLh stars, the early-mid O hypergiants also seem  to have 
evolved from very massive progenitors.

Unfortunately we are only able to find dynamical mass estimates for a handful of  O giants and supergiants, all of which are of later spectral type than found within the Arches\footnote{The primary of  V729 Cyg (O7Ianfp, $\sim 
31.9\pm3.2M_{\odot}$; Linder et al. \cite{linder}), the secondary of Wd1-13 (O9.5-B1 
Ia; $35.4{\pm}5M_{\odot}$; Ritchie et al.  \cite{ritchie}), V1007 Sco A+B 
(O7.5 III + O7 III,  $29.5{\pm}0.4 + 30.1{\pm}0.4M_{\odot}$; Mayer et al. \cite{mayer}) and CC Cas (O8.5 III, 
$35.4{\pm}5M_{\odot}$; Gies \cite{gies}).}. 

Fortunately, significantly more dynamical estimates are 
available for main sequence O stars, and in Fig. 10 we present values from Weidner \& Vink (\cite{weidner}; henceforth WeVi10) for 
galactic stars, supplemented  by more recent determinations\footnote{ V382 Cyg (O7 V + O8 V, 
$27.9{\pm}0.5 + 20.8{\pm}0.4M_{\odot}$;  Yasarsoy \& Yakut \cite{yasarsoy}), MY Cam (O4 V + O6 V, 
$37.7{\pm}1.6 + 31.6{\pm}1.4M_{\odot}$;  Lorenzo et al.  \cite{lorenzo14}) and HD 150136 (O3-3.5 
V((f$^*$)) + O5.5-6 V + O6.5-7 V((f)), $>27.7{\pm}0.4 + >17.5{\pm}0.3M_{\odot}$; Mahy et al. 
\cite{mahy}). Comparison of HD150135 to  evolutionary predictions implies $\sim64M_{\odot}$ and 
$40M_{\odot}$ respectively; both values are indicated in 
Fig. 10.}. This indicates a broad linear relationship  between mass and spectral type, with masses ranging from  
$\sim20M_{\odot}$ for O9 V stars through to $\sim30-38M_{\odot}$ for the earliest O5-6 V candidates identified within 
the Arches.

\subsubsection{Theoretical calibration of the spectral type vs. stellar mass relation}

In an effort to circumvent the lack of dynamical  mass determinations WeVi10 
and Martins \& Palacios (\cite{martins16}; henceforth MaPa17) both  utilised stellar evolutionary codes to predict the 
physical appearance of {\em single} stars of a given mass at a  particular stage of their lifecycle. The former 
authors compare the outputs of Geneva group models   (e.g. luminosity, temperature; 
Meynet \& Maeder \cite{meynet}) to the spectral  type and luminosity class calibrations of Martins et  al. (\cite{martins05}). MaPa17
adopt a different  approach, utilising  the output of the STAREVOL code (Decressin et 
al. \cite{decressin}, Amard et al. \cite{armard}) as input for the non-LTE model atmosphere code CMFGEN (Hillier \&   Miller \cite{hillier98}, \cite{hillier99}). They use CMFGEN to  generate a grid of  synthetic spectra as a function of initial stellar mass and age, which then may be subject to spectral classification utilising the same criteria as employed for observational data.

MaPa17 provide evolutionary pathways for non-rotating stars of a given {\em initial} mass. 
They show that the main sequence population within the Arches arises from stars spanning $\sim15M_{\odot}$ (O9 V)
to $\sim30-40M_{\odot}$ (O5-6 V). Stars with intermediate luminosity classes (II-IV) have initial masses ranging from 
$\sim40-50M_{\odot}$ (O6.5-7) to $\sim50-60M_{\odot}$ (O5-6). Finally the O4-5 Ia stars appear to evolve from very massive 
$\geq80M_{\odot}$ progenitors, while the handful of later O5-6.5 Ia stars are potentially consistent with an extension of
this range to $\sim60M_{\odot}$ and above.

In contrast WeVi10 provide initial {\em and} current masses for both non-rotating and rotating stars of a given spectral type and 
luminosity class. The results for O9V stars - $\sim14-24M_{\odot}$ - are broadly compatible with the findings of MaPa17.
However WeVi10 report a much wider mass range for  O5-6 V stars; $\sim26-53M_{\odot} (28-53M_{\odot})$ 
for non-rotating(rotating) models. 
This trend continues through to supergiants, with WeVi10 
suggesting that O4-5 Ia  stars may derive from progenitors with masses of $58-120M_{\odot} (52-97M_{\odot})$ for 
non-rotating (rotating) 
models, with {\em current}  masses spannning $\sim52-103M_{\odot} (43-74M_{\odot})$ respectively. Such ranges differ significantly 
from the results of MaPa17 and, if confirmed, would appear to  preclude the assignment of {\em unique} stellar masses based on spectral classification alone, especially for the earliest spectral types.

\subsubsection{Summary - stellar masses for Arches members}

Given the notable differences between the two theoretical studies summarised above, we assign primacy to dynamical mass estimates 
(Sect. 5.1.1). In doing so we  suggest that both initial and current masses  range from $\sim15-20M_{\odot}$ for the 
O8-9 V stars  within 
the Arches, through to $\sim30-40M_{\odot}$ for the O5-6 V stars. These estimates are broadly consistent with the results of MaPa17.
The {\em current} masses of the WN8-9h primary and O5-6 Ia$^+$ secondary of F2 are consistent with  other dynamical
estimates for stars of comparable spectral type suggesting that the respective cluster cohorts are likewise
of very high mass ($\sim80M_{\odot}$ and $\sim60M_{\odot}$ respectively), noting that neither theoretical study encompasses
 such stars. Finally the lack of dynamical estimates  and  the wide  mass ranges 
suggested by WeVi10 and, to a lesser extent,by MaPa17,  preclude the emplacement of robust values for the 
{\em current} masses of the Arches giant and supergiant populations, but  one might reasonably expect them to lie  between 
the preceding extremes (i.e. $\sim40-60M_{\odot}$). 

It has long been established that the stellar winds of   very massive stars significantly reduce their masses as they evolve away from 
the zero-age main sequence. This is evident in the Arches, where
comparison to evolutionary calculations suggests that the WN8-9h primary of F2 was likely born with a mass of 
$\geq120M_{\odot}$  (Lohr et al. Paper II). Given that  correction for this effect required an accurate determination of 
the current stellar parameters of F2, we refrain from  inferring {\em initial} masses for the 
more evolved stellar cohorts (the remaining WNLh and O stars of luminosity classes I-III) within the Arches at this 
time. 

\subsection{Cluster age}

\subsubsection{Comparison to other clusters}

 While determining a cluster age via isochrone fitting is impossible at 
 this juncture, we may utilise the presence - and absence - of stars of particular 
 spectral types/luminosity classes to provide a qualitative estimate. 
 The simplest approach is to compare the stellar 
content of the Arches to clusters  with more accurate age estimates due to 
lower interstellar extinction (cf. 
the compilation by Clark et al.  \cite{clark13}).  Trivially, the lack 
of cool super/hypergiants suggests  an age of $<5$ Myr 
(Clark et al. \cite{clark05}). Fortuitously, despite suffering from a similarly  uncertain extinction, a more stringent 
constraint is provided by  the Quintuplet cluster,
which hosts a substantial $\sim$O7-B0 supergiant population  - while lacking the O4-6 supergiants found
in the Arches - as well as a wealth of  early B hypergiants and  LBVs (Figer  et al. \cite{figer99b}, Liermann et al. 
\cite{liermann09}, Clark et al. in prep.).
These populations are absent from the Arches, suggesting that it is a younger system. Comparison
to the studies of  Groh et al. (\cite{groh14}) and  MaPa17 suggests an age of
 $\sim3-3.6$ Myr for the Quintuplet, which consequently would form an upper limit to the age of the Arches.

A  lower bound to the age of the Arches is suggested by the 
apparent lack of stars with spectral type O2-3  of any 
luminosity class. Specifically, 
NGC3603\footnote{Hosting six O3V, six O3 III, one O3.5Ifa, one O3If$^*$/WN6 
and three WN6ha stars for an assumed age of $\sim1-2$ Myr
 (Drissen et al. \cite{drissen}, Melena et al. \cite{melena}).}, the 
 apparently single aggregate comprising Trumpler 16 and Collinder 
 228\footnote{Hosting two O3.5 V((f)) stars and three WN6-7ha stars for an 
assumed age of $\sim2-3$ Myr (Smith \cite{smith06}).},  Trumpler 
14\footnote{Hosting one O2If$^*$ and three O3 V stars for an
assumed age of $\sim1-2$ Myr (Smith \cite{smith06}).} and R136
at the heart of  the LMC star-forming region 30 Doradus\footnote{Hosting
three WN5h, one O2 If$^*$/WN5, two O2 If, two O2 III-If, two O2-3III and 
8 O2-3 V stars for an age of $1.5^{+0.3}_{-0.7}$ Myr (Crowther et al. 
\cite{crowther16}).} all appear demonstrably younger than the Arches.

Nevertheless we are able to identify comparators, specifically 
 Danks 1 (1.5$^{+1.5}_{-0.5}$ Myr; Davies et al. \cite{davies}) and 
potentially the lower mass aggregate Havlen 
Moffat 1 ($\sim$1.7-4 Myr; Massey et al. \cite{massey}, V\'{a}zquez \& 
Baume \cite{vazquez}). Consequently such an approach yields a qualitative
age determination broadly consistent with previous quantitative estimates, 
although critically not of sufficient precision to determine whether one would expect 
SNe to have already occurred.

\subsubsection{Comparison to theoretical predictions}

 A related  approach is to use the presence - or absence - of particular  
 combinations of spectral types and luminosity classes in conjunction with
  theoretical predictions to infer ages for such stars. Trivially, such
a methodology is  susceptible to the same  uncertainties in the evolutionary
physics  that afflicted the (analogous) determination of  stellar
masses  in Sect. 5.1.2.
Moreover the ingress and egress of cluster stars from a particular evolutionary
phase is expected to be dependent on the distribution of stellar rotational velocities, which is
unconstrained for the Arches.

Mindful of these {\em caveats}, and subject to uncertainties in the spectral classification,
an obvious starting point is to utilise the location of the main sequence turn-off;  given the 
possibility that very massive objects,  such as the  WNLha stars, may still be H-burning, we use 
 this term with reference to a departure from luminosity class V.
WeVi10 demonstrate that  the apparent absence of main sequence stars
of spectral class O4 and earlier implies  a minimum age of $\sim2$ Myr for the Arches for both rotating and
 non-rotating models. Conversely, WeVi10 suggest that O5 V (O6 V) stars evolve away from the MS  after 2.4 (3.3) Myr; given their 
presence within the Arches this provides an upper limit to the cluster age. 

Are the properties of the O supergiant and giant cohorts consistent with a cluster age in the range of 2-3.3 Myr?
Turning first to  the supergiants, and such a lower limit would be consistent with the absence of O3 I-III stars, which WeVi10 suggest disappear after 1.7 Myr. They further suggest that while non-rotating O4 supergiants
vanish after $\sim2$ Myr, rotating examples persist until 2.8 Myr, while  non-rotating (rotating) O5 Ia stars may be expected until 2.5 Myr (3.4 Myr); both consonant with the age range inferred from the main sequence turn-off. Non rotating supergiants of spectral type O6 (6.5) first appear  after 2.1 (2.2) Myr and so are also expected within this window.
Thus, depending on the distribution of rotational velocities and given the uncertainties in spectral classification, the 
properties of the supergiant cohort are indeed consistent with an age of $\sim2.0-3.3$ Myr.

Interpreting the prospective giant population is more difficult given current observational uncertainties. Broadly speaking, the time of the first appearance of giants  with spectral classifications earlier than  
$\sim$O6  is consistent with the upper limit(s) to the cluster age; however objects with  spectral types later than $\sim$O6.5-7 
might be expected to appear at later times.  Consequently if stars such as F85 (O7-8 III-V) are found to be giants they may be 
in tension with the constraints implied by the main sequence population. 

Unfortunately, no theoretical predictions for the 
ages of the hypergiant and WNL populations are available to test their compatibility with these estimates.

We may, however,  utilise  the apparent lack of WC (or other hydrogen-free) stars, LBVs and B-type hypergiants within the Arches 
to provide additional constraints (Sect. 3.2.4). The  absence of WC stars implies that cluster members have yet to 
reach this evolutionary stage and, since it is thought to almost immediately precede core-collapse, it appears likely that  
SNe also have yet to occur. Groh et al. (\cite{groh13}) provide simulations of massive stars that yield  the  time  at which 
SNe take place as a function of initial stellar mass.  For $120M_{\odot}$ stars - implied
by the primary of F2 (Sect. 5.1.1) -  these suggest that SNe first occur between 3 and 3.55Myr (for non-rotating and rotating models); if the above assertions are correct the latter serves as an upper limit 
to the cluster age\footnote{Unfortunately Groh et al. (\cite{groh13}) do not provide timings for the onset of the WC 
phase for stars  $\geq85M_{\odot}$ and so we cannot allow for its duration in our age determinations, although we note that 
for  $\geq60M_{\odot}$ stars Groh et al.(\cite{groh14}) show the WC phase persists for $\sim0.2$ Myr prior to SN.}.

In a related study Groh et al. (\cite{groh14}) provide the full evolutionary sequence for a non-rotating $60M_{\odot}$ star, 
showing that a combined B-type hypergiant/LBV  phase is encountered between $\sim3.25-3.56$ Myr ($\sim10$\% of the lifetime of 
the preceding O supergiant phase). Since we may confidently  expect stars of $60M_{\odot}$ to have formed within the Arches 
(Sect. 5.1) we can assume that they  have yet to evolve  this far and consequently that the cluster is likely to be 
younger than $\sim3.2$ Myr.

To summarise - we infer a current age for the Arches of $\sim2-3.3$ Myr from the apparent position of the main sequence
turn-off. As far as we may determine, such a  value is consistent with (i) the age of $\sim2.6^{+0.4}_{-0.2}$ Myr obtained for the eclipsing SB2 binary F2 (Paper II), (ii) the  properties of the more evolved 
stars within the Arches - such as the supergiant cohort, (iii) the lack of LBVs and H-free Wolf-Rayets such as the WC and WO sub-types and (iv) the ages determined for other young massive clusters with similar  stellar populations.  Within the 
uncertainties it is also consistent with previously published  determinations, albeit lying at the lower end of 
the resultant span of age estimates.

 Furthermore given the absence of an H-free stellar cohort we consider it unlikely that 
the cluster is at an 
age at which SNe are regularly occurring, although we cannot exclude the possibility that a handful of unusually massive 
($>>120M_{\odot}$) stars have already been lost to such events.

\section{Discussion}

As previously discussed the most noteworthy features of our spectroscopic dataset appear to be the 
continuous and  smooth progression of  spectral morphologies from intrinsically luminous to less luminous cluster members and 
the  absence of any  star more evolved than the WNLh cohort. To these we may add the {\em apparent} lack of massive blue 
stragglers within the cluster.  Specifically no stars with spectral types O2-3, such as those seen in e.g. NGC3603 and R136, are 
present (footnotes 16 and 19 and Crowther \& Walborn \cite{crowther11}). Likewise no WN5-6h stars are found within the 
Arches, 
although they have been identified in younger clusters such as NGC3603 (Melena et al 
\cite{melena}).

What constraints do such observations place on stellar evolution? The  distribution of spectral types amongst 
the candidate main sequence and giant stars is consistent with single star evolution {\em and} the simulations of 
Schneider et al. (\cite{schneider14})  which imply that the majority of stars in the $\sim32-50M_{\odot}$ window 
are pre-interaction systems. However the situation differs for  
supergiants and 
more 
luminous/evolved  stars.  Ma08 interpret the distribution of spectral types  under the paradigm of single star evolution, 
suggesting that they are consistent with an evolutionary sequence progressing from O supergiant through O hypergiant to WNha 
star (for masses $>60M_{\odot}$), with the latter stars  evolving from earlier (O2-3)  supergiants than are currently 
present within 
the Arches.  Conversely, Schneider et al. (\cite{schneider14}) suggest - under the assumption of a 100\% initial binary 
fraction -  that the WNha stars are instead binary products.

Two observational findings cast light on these assertions. Firstly there is a smooth    
progression from O supergiants with infilled Br$\gamma$ absorption (e.g. F20, F33 and B4; Fig. 3) through the least 
extreme hypergiants (e.g. F18, F27 and F40; Fig. 2) to those with the strongest emission in Br$\gamma$ (F10, F17 and the 
secondary in F2; Fig. 2) which in turn are almost indistinguishable  from   WNLha stars such as F16 (Fig. 1). The 
development of increased 
emission in Br$\gamma$ is indicative of a progressively  denser stellar wind across these spectral types and reinforces  the
supposition that the hypergiants indeed bridge the gap between the O supergiant and WNLha stars.
Secondly the physical properties of the eclipsing binary F2 suggest it is in a pre-interaction phase (Paper II). Analysis of its lightcurve suggests that it is just entering a contact phase with the orbit found to  be slightly 
eccentric. Moreover the current WN8-9h primary of F2 appears in a more evolved phase than the  O4-5Ia$^+$ secondary; if it 
were the product of mass transfer the reverse would be the case.

As a consequence we strongly suspect that at least some of the WNLh cohort are indeed the product of the  single star evolutionary 
channel proposed by Ma08. However the presence of F2 and the hard X-ray emission from a number of other WNh stars is clearly 
indicative of a binary population amongst these objects (Wang et al. \cite{wang}). Consequently one might ask where the post-interaction binaries are given that 
Schneider et al. (\cite{schneider14}) suggest that {\em on average} the most massive star in an Arches-like cluster would be expected to be a binary product after only $\sim1$ Myr?

Obvious routes to lower the expected number of  blue stragglers would  be to reduce one or more of the cluster mass, age or binary fraction and/or to steepen the initial mass function. 
However if exceptionally massive stars always present as WN7-9h in high metallicity environments such as the galactic centre, rather than the WN5-6h objects seen   in e.g. NGC3603 and R136,  once could suppose that blue stragglers are hidden in plain sight amongst the most luminous examples of this population. 

\section{Conclusions}

In this paper we present the first results of a multi-epoch spectroscopic survey of the Arches cluster, co-adding 
multiple spectra to  obtain the deepest observations ever of the stellar population. We  supplement these with 
new HST photometric data for confirmed and candidate cluster members. Excluding interlopers and those objects with  low S/N 
and/or blended spectra, we provide spectral classifications for 88 cluster members, an increase in sample size over Ma08 
by a factor of $\sim3$.  We find no further WRs of any  subtype in the cluster; importantly, no H-free stars have been identified. In contrast we expand the number of cluster O hypergiants from two to eight and  
supergiants from  11 to 25; the largest population of any known Galactic cluster. 
The greater S/N of these data allow us to refine previous classifications but, 
as with Ma08,  no examples of supergiants with  spectral type earlier than O4 or later than O6.5 are found.
Extending the sequence of morphologically similar spectra to fainter objects, we are able to identify 
a population of intermediate (I-III) luminosity class stars which smoothly segues into a  cohort of  
giants with spectral types covering the same range as the supergiants. Finally we identify a  
number of fainter objects which  we classify as main sequence stars, with spectral types 
ranging from O5-6 V to $\geq$O8 V.  This implies a main-sequence turn-off between O4-5V (where we employ this term to 
refer specifically to luminosity class V objects, since it is suspected that some evolved, very massive stars may still
be core H-burning;  e.g. Smith \& Conti \cite{smith08}).

Provisional analysis of a number of the WNLh stars reveals that a combination of uncertainty in the correct extinction law to apply 
and differential reddening across the cluster leads to unexpectedly large errors in 
 luminosity ($\Delta L_{\rm bol} \sim0.6$dex and $\sim0.4$dex, respectively). 
While our results favour a Moneti rather than a power-law formulation for the extinction law, accurate 
determination of the stellar parameters of cluster members requires modelling of individual objects, beyond the scope of 
this work. As such we are not able to make direct comparison to theoretical isochrones to determine a cluster age, calibrate the 
(I)MF or determine an integrated cluster mass, and urge caution with regard to previous determinations - and conclusion resulting from them - 
given that uncertainties employed in their calculation  have been systematically underestimated.

Nevertheless we are able to estimate stellar and cluster properties from the current data. Specifically the dynamical masses of 
the WN8-9h+O4-6Ia$^+$ binary F2 (Paper II) suggest current masses of $\sim80M_{\odot}$ and $\sim60M_{\odot}$ for the WNLh 
and O hypergiant cohorts respectively. Comparison of the stellar properties of the WN8-9h primary to evolutionary tracks 
suggests an initial mass of $\geq120M_{\odot}$ (Paper II), providing strong support for the modification of the upper reaches of the current mass function due to  mass-loss (cf. Schneider et al. \cite{schneider14}). Empirical calibration of the spectral  type/mass relation for main sequence stars suggests a main-sequence turn-off around $\sim30-38M_{\odot}$. Consequently, excluding the 13 stars which we classify as 
$\geq$O8 V, the Arches appears to contain at least 75 stars with initial masses $\gtrsim30M_{\odot}$, with the masses of the super-/hypergiants and WNLh stars greatly in excess of this value.

The main sequence turn-off suggests a cluster age of $\sim2-3.3$ Myr, broadly consistent with the properties of the 
more evolved population of the cluster, including the eclipsing binary F2 (Paper II). Notably, the lack of H-free WRs, BHGs and LBVs  strongly argues against ages much larger than this. Comparison to the Quintuplet also suggest an upper limit to the  age of the Arches of  $\lesssim3.6$ Myr (Clark et al. in prep.), consistent with the above results.
Unfortunately these estimates still bracket the ages at which one might expect the first SNe to occur for very 
massive stars ($\geq120M_{\odot}$). An expanded sample of high S/N spectra of main sequence stars would  help refine 
this critical parameter.

 The smooth evolution in spectral morphologies transiting from O supergiant through the greatly expanded hypergiant population to 
the WNLh stars, when combined with the properties of the apparently pre-interaction binary F2, argues for the prevalence of a 
single star evolutionary  pathway within the  cluster at this time. Nevertheless  the presence of a number of very massive binaries 
(Sect. 6 and Paper III) suggests that a binary channel may well play an important  (future) role. Despite predictions that binary 
products may be present   within the cluster after only $\sim1$ Myr we see no systems currently exhibiting mass-transfer nor blue stragglers, unless both 
single and binary evolution yield superficially identical WNLh stars.

In conclusion the combination of the relative proximity, age and the  rich stellar population of the Arches makes it a unique 
laboratory for studying the evolution of very massive stars via both single and binary evolutionary channels as well as star 
formation in the extreme environment of the Galactic centre. In particular we may hope to constrain the upper mass 
limit for  very massive stars, determine how they form (and/or how they subsequently grow via binary interaction)  
and elucidate the physical properties of this population. All of these goals are essential  for quantifying chemical, 
mechanical 
and radiative feedback in both the local and early Universe; moreover, such stars are the prime candidates for 
pair-instability SNe, GRBs and ultimately coalescing binary black holes. However in order to fulfil the potential of the Arches 
in these regards we must first constrain the properties of the binary population and the physical parameters of the 
constituent  stars via individual quantitative modelling: goals that will be addressed in future papers in this 
series.

\begin{acknowledgements}
Based on observations collected at the European Organisation for Astronomical Research in the 
Southern Hemisphere under ESO programmes  087.D-0317,  091.D-0187 and
099.D-0345. This research was supported by the Science and
Technology Facilities Council. FN acknowledges financial support through Spanish grants 
ESP2015-65597-C4-1-R and ESP2017-86582-C4-1-R (MINECO/FEDER). We thank Chris Evans and Paul Crowther for their valuable comments.
Finally we thank the referee, Ben Davies, for his careful reading and insightful comments, which significantly improved the manuscript.

\end{acknowledgements}

{}

\appendix

\section{Online Material}

\longtab{1}{
\begin{longtable}{lccccccccl}
\caption{The stellar population of the Arches cluster}\\
\hline
\hline
ID &        RA     &       Dec     &  $m_{\rm F127M}$   & $m_{\rm F139M}$   &   $m_{\rm F153M}$  & $m_{\rm F205W}$ & \#Observations & 
Spec. & Notes \\
   &    (h m s)    &     (d m s)   &     (mag)  & (mag)     &   (mag)       &  (mag)     & (\#Epochs)   & Class. &       \\
\hline
\endfirsthead
\caption{continued.}\\
\hline
\hline
ID &        RA     &       Dec     &  $m_{\rm F127M}$   & $m_{\rm F139M}$   &   $m_{\rm F153M}$  & $m_{\rm F205W}$ & \#Observations &
Spec. & Notes \\
   &    (h m s)    &     (d m s)   &     (mag)  & (mag)     &   (mag)       &  (mag)     & (\#Epochs)   & Class. &       \\
\hline
\endhead
\hline
\endfoot
B1 &  17 45  51.50 &  -28 49 26.8  &       -    &   -        &     -        &    -         &     5(5)  & WN8-9h &        
\\
B4 &  17 45  50.86 &  -28 49 19.7  &       -    &   -        &     -        &    -         &    15(10)  & O5.5-6 Ia&     \\    
F1 &  17 45 50.260 &  -28 49 22.76 &  15.45     &  14.21     &   12.97    & 10.45 & 10(9) &  WN8-9h & radio$^1$ \\
F2 &  17 45 49.746 &  -28 49 26.29 &  16.73     &  15.29     &   13.94    & 11.18 & 15(14) & WN8-9h  & SB2, radio, X-ray \\ 
   &               &               &            &            &            &       &        & O5-6 Ia$^+$ & \\
F3 &  17 45 50.884 &  -28 49 26.89 &  15.05     &  13.94     &   12.81    & 10.46 & 5(5) &  WN8-9h &  radio \\
F4 &  17 45 50.628 &  -28 49 18.10 &  14.57     &  13.56     &   12.56    & 10.37 & 26(10) & WN7-8h  & radio \\
F5 &  17 45 50.510 &  -28 49 32.40 &  15.68     &  14.50     &   13.32    & 10.86 & 5(5) &  WN8-9h & radio$^1$ \\ 
F6 &  17 45 50.478 &  -28 49 22.79 &  14.60     &  13.49     &   12.39    & 10.37 & 13(5) &  WN8-9h &  radio$^1$, X-ray \\
F7 &  17 45 50.529 &  -28 49 20.03 &  14.71     &  13.63     &   12.58    & 10.48 & 26(10) &  WN8-9h &  X-ray \\
F8 &  17 45 50.447 &  -28 49 21.75 &  15.12     &  14.03     &   12.91    & 10.76 & 8(5) &  WN8-9h & radio \\
F9 &  17 45 50.321 &  -28 49 12.26 &  14.91     &  13.85     &   12.82    & 10.77 & 10(10) &  WN8-9h &  X-ray \\
F10&  17 45 50.121 &  -28 49 27.01 &  16.21     &  14.93     &   13.74    & 11.46 & 5(5) &  O7-8 Ia$^+$ &  \\ 
F12&  17 45 50.337 &  -28 49 17.78 &  15.35     &  14.26     &   13.21    & 10.99 & 13(10)$^b$ & WN7-8h &  \\
F13&  17 45 50.102 &  -28 49 24.15 &  16.52     &  15.28     &   14.08    & 11.74 & 5(5) & O7-8 Ia$^+$ &  \\
F14&  17 45 50.735 &  -28 49 23.08 &  15.43     &  14.34     &   13.31    & 11.72 & 12(5) & WN8-9h &  \\
F15&  17 45 50.811 &  -28 49 17.09 &  15.07     &  14.05     &   13.07    & 11.27 & 20(11) & O6-7 Ia$^+$ &  \\
F16&  17 45 50.581 &  -28 49 21.17 &  15.55     &  14.48     &   13.42    & 11.40 & 11(5) & WN8-9h &  \\  
F17&  17 45 50.192 &  -28 49 27.66 &  17.03     &  15.73     &   14.47    & 12.15 & 5(5) &  O5-6 Ia$^+$ &  \\
F18&  17 45 50.532 &  -28 49 18.42 &  15.63     &  14.59     &   13.61    & 11.63 & 23(10) &  O4-5 Ia$^+$&  radio\\
F19&  17 45 49.818 &  -28 49 26.48 &  17.92     &  16.53     &   15.18    & 12.60 & 14(13) & O4-5 Ia &  radio \\
F20&  17 45 50.481 &  -28 49 20.18 &  16.53     &  15.41     &   14.38    & 12.16 & 20(10) &  O4-5 Ia &  \\
F21&  17 45 50.820 &  -28 49 20.11 &  15.72     &  14.67     &   13.68    & 11.77 & 22(10) & O6-6.5 Ia &  \\
F22&  17 45 50.278 &  -28 49 17.21 &  16.42     &  15.26     &   14.16    & 12.02 & 9(9)$^a$ &  O5.5-6 Ia &  \\
F23&  17 45 51.211 &  -28 49 23.84 &  16.38     &  15.27     &   14.21    & 12.19 & 1(1) &  O6-6.5 Ia &  \\
F24&  17 45 50.152 &  -28 49 21.21 &  17.29     &  16.06     &   14.88    & 12.61 & 6(6)$^c$ & O4-5 Ia &  \\
F25&  17 45 50.012 &  -28 49 27.06 &  18.35     &  16.93     &   15.57    & 13.05 & 9(9)$^b$ & O4-5 Ia &  \\
F26&  17 45 50.610 &  -28 49 24.03 &  16.61     &  15.46     &   14.37    & 12.34 & 8(5) &  O4-5 Ia&  \\
F27&  17 45 50.664 &  -28 49 20.02 &  16.09     &  15.01     &   13.97    & 12.01 & 29(10) & O4-5 Ia$^+$ &  \\
F28&  17 45 50.699 &  -28 49 22.21 &  16.17     &  15.07     &   14.06    & 12.17 & 13(5) & O4-5 Ia &  \\ 
F29&  17 45 50.799 &  -28 49 18.14 &  16.14     &  15.10     &   14.10    & 12.26 & 22(11) & O5.5-6 Ia &  \\
F30&  17 45 50.275 &  -28 49 19.10 &  16.75     &  15.63     &   14.56    & 12.53 & 10(8)$^c$ & O4-5 Ia &  \\
F31&  17 45 50.478 &  -28 49 20.16 &    -       &    -       &     -      & 12.41 & - & - &  \\
F32&  17 45 50.681 &  -28 49 20.35 &  16.46     &  15.41     &   14.37    & 12.42 & 13(5) & O4-5 Ia &  \\
F33&  17 45 50.723 &  -28 49 20.40 &  16.38     &  15.32     &   14.32    & 12.42 & 13(5) &  O4-5 Ia&  \\ 
F34&  17 45 50.863 &  -28 49 21.54 &  16.54     &  15.48     &   14.44    & 12.49 & 6(5)$^b$ &  O4-5 Ia &  \\
F35&  17 45 50.755 &  -28 49 17.50 &  16.17     &  15.15     &   14.18    & 12.37 & 20(11) & O4-5 Ia &  \\
F36&  17 45 49.789 &  -28 49 07.89 &  18.03     &  16.58     &   15.13    & 12.60 & - & - &  \\
F37&  17 45 50.529 &  -28 49 19.77 &    -       &    -       &     -      & 12.63 & - & - &  radio$^1$ \\
F38&  17 45 50.755 &  -28 49 20.48 &  16.33     &  15.36     &   14.22    & 12.38 & 11(5) &  O4-5 Ia &  \\
F39&  17 45 51.166 &  -28 49 36.70 &  16.98     &  15.84     &   14.76    & 12.65 & - & - &  \\
F40&  17 45 50.685 &  -28 49 18.83 &  16.63     &  15.57     &   14.55    & 12.67 & 27(11) & O4-5 Ia$^+$ &  \\
F41&  17 45 50.429 &  -28 49 27.42 &  18.46     &  17.15     &   15.89    & 13.53 & - & - & \\
F42&  17 45 50.364 &  -28 49 21.24 &  16.99     &  15.89     &   14.85    & 12.82 & 10(8) &  O4-5 Ia &  \\
F43&  17 45 50.546 &  -28 49 19.17 &  17.28     &  16.21     &   15.19    & 13.04 & 25(11) & O4-5 Ia &  \\
F44&  17 45 50.701 &  -28 49 25.59 &  17.19     &  16.05     &   14.93    & 12.88 & 5(3) & -  & blend \\
F45&  17 45 50.497 &  -28 49 24.25 &  17.33     &  16.16     &   15.07    & 13.01 & 10(5) & O4-5 Ia &  \\
F47&  17 45 50.754 &  -28 49 25.28 &  17.02     &  15.92     &   14.87    & 12.90 & 8(6) & O4-5 Ia &  \\
F48&  17 45 50.623 &  -28 49 26.96 &  17.69     &  16.53     &   15.42    & 13.28 & - & - &  \\
F49&  17 45 50.128 &  -28 49 07.79 &  17.42     &  16.25     &   15.12    & 12.94 & 3(3) & O5.5-6 Ia   &  \\
F50&  17 45 50.134 &  -28 49 26.17 &  18.30     &  17.00     &   15.77    & 13.53 & 6(5) & O4-5 Ia  &  \\
F52&  17 45 51.206 &  -28 49 33.31 &  17.64     &  16.41     &   15.24    & 12.94 & - & - &  \\
F53&  17 45 50.714 &  -28 49 25.01 &  17.11     &  15.97     &   14.93    & 12.94 & 5(3) & O4-5 Ia &  \\
F54&  17 45 50.271 &  -28 49 15.81 &  17.93     &  16.63     &   15.42    & 13.02 & 6(6) & O5.5-6 I-III  &  \\
F55&  17 45 50.787 &  -28 49 22.61 &  17.11     &  16.02     &   15.00    & 13.03 & 6(5) & O5.5-6 III  &  \\
F56&  17 45 50.591 &  -28 49 22.19 &  17.00     &  15.92     &   14.91    & 13.03 & - & - &  \\
F57&  17 45 50.768 &  -28 49 20.31 &  16.91     &   -        &     -      & 13.04 & - & - &  \\
F58&  17 45 49.968 &  -28 49 19.60 &  17.83     &  16.57     &   15.37    & 13.05 & - & - &  \\
F59&  17 45 51.138 &  -28 49 14.34 &  17.05     &  15.98     &   14.95    & 13.05 & - & - &  \\
F60&  17 45 50.799 &  -28 49 22.18 &  17.21     &  16.08     &   15.05    & 13.02 & 6(5) & O4-5 Ia  &  \\
F61&  17 45 50.144 &  -28 48 59.09 &  17.38     &  16.24     &   15.06    & 13.09 & - & - &  \\
F62&  17 45 50.427 &  -28 49 16.61 &  17.04     &  15.99     &   15.00    & 13.04 & 17(14) & O4-5 I-III &  \\
F63&  17 45 50.838 &  -28 49 25.68 &  17.20     &  16.15     &   15.14    & 13.15 & 7(6) & O4-5 I-III &  \\
F64&  17 45 50.481 &  -28 49 16.71 &  17.15     &  16.10     &   15.08    & 13.13 & 14(10)  & O4-5 I-III  &  \\
F65&  17 45 50.082 &  -28 49 21.53 &  17.58     &  16.42     &   15.32    & 13.16 & 1(1) & O4-5 I-III  & low S/N  \\
F66&  17 45 50.532 &  -28 49 20.42 &  17.07     &  15.99     &   15.00    & 13.11 & - & - &  \\
F67&  17 45 50.475 &  -28 49 15.10 &  17.69     &  16.56     &   15.48    & 13.35 & - & - &  \\
F68&  17 45 50.855 &  -28 49 18.15 &  16.70     &  15.68     &   14.71    & 12.93 & 14(11) & O5.5-6 III &  \\
F69&  17 45 50.847 &  -28 49 20.32 &  17.06     &  15.99     &   14.97    & 13.06 & 4(4) & O6-6.5 III  &  \\
F70&  17 45 50.252 &  -28 49 23.39 &  17.82     &  16.59     &   15.42    & 13.30 & - & - &  \\ 
F71&  17 45 50.060 &  -28 49 16.43 &  18.14     &  16.92     &   15.77    & 13.62 & - & - &  \\
F72&  17 45 50.784 &  -28 49 20.27 &    -       &  16.08     &   15.05    & 13.13 & - & - &  \\
F73&  17 45 50.679 &  -28 49 25.58 &  17.78     &  16.59     &   15.48    & 13.35 & 5(3) &- & blend \\
F74&  17 45 50.993 &  -28 49 27.49 &  17.62     &  16.50     &   15.42    & 13.36 & 5(5) & O5-6.5 I-III &  \\
F75&  17 45 50.825 &  -28 49 11.25 &  18.91     &  17.48     &   16.05    & 13.36 & - & - &  \\
F76&  17 45 50.341 &  -28 49 17.06 &  17.64     &  16.49     &   15.38    & 13.38 & - & - &  \\
F77&  17 45 50.771 &  -28 49 19.83 &  17.30     &  16.26     &   15.23    & 13.25 & 25(10) & O6-6.5 III &  \\
F78&  17 45 51.939 &  -28 49 24.71 &  17.94     &  16.77     &   15.59    & 13.44 & - & - &  \\
F79&  17 45 50.074 &  -28 49 18.11 &  19.20     &  17.72     &   16.23    & 13.46 & - & - &  \\
F80&  17 45 50.589 &  -28 49 19.46 &  17.40     &  16.29     &   15.32    & 13.47 & 24(10) & - & blend  \\
F81&  17 45 50.713 &  -28 49 24.24 &  17.62     &  16.51     &   15.46    & 13.48 & 10(5) & O6-7 III-V &  \\
F82&  17 45 50.736 &  -28 49 19.13 &  17.37     &  16.30     &   15.31    & 13.40 & 29(10) & O5-6 III-V &  \\
F83&  17 45 50.061 &  -28 49 20.67 &  18.00     &  16.81     &   15.69    & 13.53 & - & - &  \\
F84&  17 45 50.526 &  -28 49 16.28 &  17.58     &  16.49     &   15.61    & 13.54 & 18(10) & O6-7 III-V &  \\
F85&  17 45 50.484 &  -28 49 18.60 &  17.50     &  16.47     &   15.46    & 13.54 & 17(10) & O7-8 III-V &  \\
F86&  17 45 50.985 &  -28 49 33.59 &  17.30     &  16.32     &   15.35    & 13.56 & - & - &  \\
F87&  17 45 50.518 &  -28 49 16.58 &  17.62     &  16.58     &   15.56    & 13.60 & 19(10) & O5-6 V &  \\
F88&  17 45 49.653 &  -28 49 17.23 &  19.10     &  17.66     &   16.22    & 13.62 & - & - &  \\
F89&  17 45 50.955 &  -28 49 17.15 &  17.33     &  16.33     &   15.39    & 13.65 & 11(10) & O6-7 V &  \\
F90&  17 45 50.570 &  -28 49 32.97 &  18.30     &  17.09     &   15.94    & 13.68 & 5(5) & O5-6 V & Br$\gamma$ infilled  \\
F91&  17 45 49.978 &  -28 49 17.52 &  18.43     &  17.19     &   16.03    & 13.69 & - & - &  \\
F92&  17 45 50.518 &  -28 49 21.45 &  17.59     &  16.50     &   15.44    & 13.48 & 12(5) &  O5-6 V & Br$\gamma$ infilled\\
F93&  17 45 50.376 &  -28 49 22.78 &  18.14     &  17.04     &   15.99    & 13.81 & 10(8) &  O5-6 III-V & Br$\gamma$ infilled \\
F94&  17 45 50.657 &  -28 49 28.03 &  18.74     &  17.50     &   16.33    & 14.14 & - & - &  \\
F95&  17 45 50.578 &  -28 49 22.00 &  17.80     &  16.71     &   15.69    & 13.71 & - & - &  \\
F96&  17 45 50.724 &  -28 49 19.73 &  17.36     &  16.35     &   15.37    & 13.54 & 29(10) &  O5-6 III &  \\
F97&  17 45 51.385 &  -28 49 16.57 &  17.83     &  16.72     &   15.67    & 13.73 & - & - &  \\
F98&  17 45 51.153 &  -28 49 37.06 &  18.08     &  16.93     &   15.87    & 13.75 & - & - &  \\
F100& 17 45 51.061 &  -28 49 17.61 &  17.57     &  16.55     &   15.59    & 13.77 & - & - &  \\
F101& 17 45 50.914 &  -28 49 18.37 &  17.55     &  16.53     &   15.58    & 13.78 & 14(11) & O6-8 V &  \\
F102& 17 45 50.876 &  -28 49 28.80 &  18.24     &  17.07     &   15.93    & 13.78 & 4(4) & O5-8 V &  \\
F103& 17 45 49.679 &  -28 49 22.58 &  20.16     &  18.68     &   17.17    & 14.53 & - & - &  \\
F104& 17 45 50.577 &  -28 49 43.77 &  18.09     &  16.96     &   15.85    & 13.79 & - & - &  \\
F105& 17 45 50.596 &  -28 49 39.60 &  19.13     &  17.75     &   16.37    & 13.81 & - & - &  \\
F106& 17 45 49.886 &  -28 49 23.23 &    -       &   -        &     -      & 13.82 & - & - &  \\
F107& 17 45 49.872 &  -28 49 30.99 &  18.01     &  16.89     &   15.77    & 13.87 & - & - &  \\
F108& 17 45 50.809 &  -28 49 10.93 &  18.12     &  17.00     &   15.94    & 13.89 & - & - &  \\
F109& 17 45 50.520 &  -28 49 20.70 &  18.00     &  16.91     &   15.86    & 13.91 & - & - &  \\
F110& 17 45 50.574 &  -28 49 16.35 &  18.00     &  16.93     &   15.90    & 13.93 & 18(10) &$\geq$O8 V  &  \\
F111& 17 45 50.309 &  -28 49 03.86 &  18.45     &  17.25     &   16.13    & 13.94 & - & - &  \\
F112& 17 45 50.507 &  -28 49 16.36 &  17.86     &  16.80     &   15.81    & 13.87 & 14(10) & O6-8 V &  \\
F113& 17 45 50.017 &  -28 49 17.47 &  18.67     &  17.42     &   16.22    & 13.97 & - & - &  \\
F114& 17 45 50.804 &  -28 49 19.07 &  17.95     &  16.90     &   15.90    & 13.99 & 22(10) & O6-8 V &  \\
F115& 17 45 50.437 &  -28 49 20.04 &  17.99     &  16.91     &   15.83    & 13.83 & 15(10) & O5-6 V &  \\
F116& 17 45 50.537 &  -28 49 06.28 &  18.77     &  17.68     &   16.63    & 14.01 & - & - &  \\
F117& 17 45 50.755 &  -28 49 19.44 &  18.05     &  16.99     &   15.97    & 14.06 & 25(10) & O6-8 V  &  \\
F118& 17 45 50.453 &  -28 49 25.97 &  18.66     &  17.44     &   16.28    & 14.08 &  4(2) & $\geq$O8 V & low S/N \\
F119& 17 45 50.842 &  -28 49 22.75 &  18.18     &  17.09     &   16.06    & 14.06 & 6(5) &  $\geq$O8 V & low S/N \\ 
F120& 17 45 50.692 &  -28 49 29.96 &  18.75     &  17.52     &   16.34    & 14.09 & - &  -&  \\  
F121& 17 45 50.204 &  -28 49 19.64 &  18.53     &  17.35     &   16.24    & 14.09 & 5(5) &  $\geq$O8 V & low S/N \\
F122& 17 45 50.358 &  -28 49 17.28 &  18.17     &  17.09     &   16.05    & 14.09 & - & - &  \\
F123& 17 45 50.685 &  -28 49 22.35 &    -       &    -       &     -      & 14.11 & - & - &  \\
F124& 17 45 50.718 &  -28 49 17.26 &  18.48     &  17.42     &   16.38    & 14.11 & - & - &  \\
F125& 17 45 51.488 &  -28 49 21.75 &  19.57     &  18.12     &   16.69    & 14.15 & - & - &  \\
F126& 17 45 50.930 &  -28 49 03.63 &  17.93     &  16.93     &   15.98    & 14.16 & - & - &  \\
F127& 17 45 51.184 &  -28 49 31.52 &  19.45     &  18.05     &   16.67    & 14.16 & - & - &  \\
F128& 17 45 51.216 &  -28 49 36.53 &  18.51     &  17.37     &   16.28    & 14.18 & - & - &   \\
F129& 17 45 49.529 &  -28 49 12.72 &  19.73     &  18.27     &   16.82    & 14.20 & - & - &  \\
F130& 17 45 50.605 &  -28 49 27.50 &  18.80     &  17.58     &   16.40    & 14.21 & - & - &  \\
F131& 17 45 50.163 &  -28 49 27.46 &  19.11     &  17.80     &   16.55    & 14.23 & 1(1) & $\geq$O8 V & low S/N \\
F132& 17 45 50.796 &  -28 49 02.68 &  18.47     &  17.34     &   16.28    & 14.27 & - & - &  \\
F133& 17 45 50.771 &  -28 49 14.97 &  18.24     &  17.19     &   16.19    & 14.27 & - & - &  \\
F134& 17 45 50.499 &  -28 49 36.54 &  19.21     &  17.93     &   16.75    & 14.28 & - & - &  \\
F135& 17 45 50.363 &  -28 49 25.73 &  18.83     &  17.61     &   16.43    & 14.29 & 2(2)  & $\geq$O8 V & low S/N \\
F136& 17 45 50.077 &  -28 49 27.84 &  19.59     &  18.18     &   16.83    & 14.30 & 5(5) & O6-8 V &  \\
F137& 17 45 51.486 &  -28 49 21.26 &  19.94     &  18.45     &   16.93    & 14.32 & - & - &  \\
F138& 17 45 50.212 &  -28 49 39.21 &  19.64     &  18.24     &   16.90    & 14.35 & - & - &  \\
F139& 17 45 51.207 &  -28 49 23.41 &  18.59     &  17.48     &   16.33    & 14.36 & 1(1) & - & low S/N \\
F140& 17 45 50.564 &  -28 49 18.48 &  18.44     &  17.38     &   16.34    & 14.38 & - & - &  \\
F141& 17 45 48.827 &  -28 49 18.81 &  20.08     &  18.57     &   17.06    & 14.39 & - & - &  \\
F142& 17 45 50.689 &  -28 49 24.93 &  18.63     &  17.50     &   16.43    & 13.41 & - & - &  \\
F143& 17 45 50.716 &  -28 49 17.24 &    -       &    -       &     -       & 14.43 & - & - &  \\
F144& 17 45 51.526 &  -28 49 20.58 &  19.33     &  18.05     &   16.79    & 14.43 & - & - &  \\
F145& 17 45 51.445 &  -28 49 20.07 &  18.45     &  17.41     &   16.38    & 14.44 & - & - &  \\
F146& 17 45 49.867 &  -28 49 14.90 &  18.88     &  17.70     &   16.57    & 14.45 & - & - &  \\
F147& 17 45 51.354 &  -28 49 20.00 &  18.60     &  17.52     &   16.47    & 14.46 & - & - &  \\
F148& 17 45 51.785 &  -28 49 28.42 &  18.69     &  17.55     &   16.53    & 14.47 & - & - &  \\
F149& 17 45 50.682 &  -28 49 01.56 &  19.49     &  18.20     &   16.86    & 14.47 & - & - &  \\
F150& 17 45 50.456 &  -28 49 19.64 &  18.50     &  17.38     &   16.34    & 14.31 & 18(9) & - & featureless? \\
F151& 17 45 50.829 &  -28 49 27.63 &  18.92     &  17.75     &   16.62    & 14.48 & 5(5) & - & featureless \\
F152& 17 45 50.695 &  -28 49 36.56 &  19.21     &  17.96     &   16.77    & 14.48 & - & - &  \\
F153& 17 45 50.360 &  -28 49 20.00 &  19.15     &  17.92     &   16.74    & 14.51 & 8(6) & $\geq$O8 V & low S/N \\
F154& 17 45 49.992 &  -28 49 33.20 &  19.45     &  18.15     &   16.89    & 14.52 & - & - &  \\
F155& 17 45 50.422 &  -28 49 19.13 &  18.65     &  17.57     &   16.53    & 14.52 & 13(9) & O6-8 V &  \\
F156& 17 45 50.642 &  -28 49 02.15 &  19.31     &  18.04     &   16.76    & 14.54 & - & - &  \\
F157& 17 45 50.220 &  -28 49 24.18 &  19.11     &  17.89     &   16.71    & 14.55 & 10(10) & $\geq$O8 V & low S/N \\
F158& 17 45 51.096 &  -28 49 21.99 &  19.09     &  17.91     &   16.80    & 14.56 & - & - &  \\
F159& 17 45 50.882 &  -28 49 15.69 &  19.63     &  18.49     &   17.32    & 14.57 & 1(1) & - &  low S/N \\
F160& 17 45 50.500 &  -28 49 15.81 &  18.61     &  17.55     &   16.55    & 14.59 & - & - &  \\
F161& 17 45 50.895 &  -28 49 40.68 &  19.75     &  18.43     &   17.06    & 14.59 & - & - &  \\
F162& 17 45 50.820 &  -28 49 39.10 &  18.85     &  17.69     &   16.56    & 14.59 & - & - &  \\
F163& 17 45 50.436 &  -28 49 23.40 &  19.84     &    -       &     -      & 14.60 & - & - &  \\
F164& 17 45 50.862 &  -28 49 08.47 &  18.60     &  17.54     &   16.54    & 14.61 & - & - &  \\
F165& 17 45 51.160 &  -28 49 15.36 &  18.60     &  17.50     &   16.52    & 14.65 & - & - &  \\
F166& 17 45 50.675 &  -28 49 19.57 &  18.62     &  17.57     &   16.52    & 14.69 & 23(9) & - & blend \\
F167& 17 45 50.733 &  -28 49 17.73 &  18.24     &  17.22     &   16.29    & 14.42 & - & - &  \\
F168& 17 45 50.495 &  -28 49 30.21 &  19.47     &  18.21     &   17.00    & 14.70 & 1(1) & - & low S/N \\  
F169& 17 45 50.377 &  -28 49 20.26 &  19.13     &  17.91     &   16.84    & 14.72 & - & - &  \\
F170& 17 45 50.704 &  -28 49 20.96 &  18.74     &  17.67     &   16.63    & 14.74 & 12(5) & - & blend \\
F171& 17 45 49.275 &  -28 49 12.13 &  19.19     &  18.00     &   16.86    & 14.76 & - & - &  \\
F172& 17 45 50.232 &  -28 49 17.52 &  19.05     &  17.90     &   16.84    & 14.76 & 4(4) & $\geq$O8 V & low S/N \\
F173& 17 45 50.204 &  -28 49 20.24 &  19.10     &  17.93     &   16.83    & 14.80 & 1(1) & - & low S/N \\
F174& 17 45 50.745 &  -28 49 28.20 &  19.26     &  18.09     &   16.95    & 14.81 & 3(3) & - & low S/N  \\
F175& 17 45 50.463 &  -28 49 03.29 &  19.72     &  18.44     &   17.18    & 14.82 & - & - &  \\
F176& 17 45 50.283 &  -28 49 23.85 &  19.25     &  18.05     &   16.94    & 14.83 &  3(3) & - &  featureless\\
F177& 17 45 50.604 &  -28 49 16.87 &  18.94     &  17.85     &   16.81    & 14.83 & 12(8) &  $\geq$O8 V & low S/N \\
F178& 17 45 50.824 &  -28 49 06.95 &  20.09     &  18.72     &   17.32    & 14.85 & - & - &  \\
F179& 17 45 51.356 &  -28 49 21.37 &  19.53     &  18.30     &   17.08    & 14.85 & - & - &  \\
F180& 17 45 50.284 &  -28 48 59.64 &  19.72     &  18.43     &   17.15    & 14.89 & - & - &  \\
F181& 17 45 50.567 &  -28 49 07.70 &  18.84     &  17.79     &   16.79    & 14.90 & - & - &  \\
F182& 17 45 50.586 &  -28 49 22.91 &  19.00     &  17.85     &   16.76    & 14.70 & - & - &  \\
F183& 17 45 49.872 &  -28 49 42.31 &  20.25     &  18.83     &   17.39    & 14.93 & - & - &  \\
F184& 17 45 50.008 &  -28 49 29.43 &  20.15     &  18.75     &   17.40    & 14.94 & 3(3) & - & low S/N \\
F185& 17 45 50.282 &  -28 49 27.56 &  19.65     &  18.40     &   17.18    & 14.95 & 2(2) &  $\geq$O8 V & low S/N \\
F186& 17 45 50.001 &  -28 49 28.24 &  20.43     &  19.01     &   17.61    & 14.95 & 6(6) &  $\geq$O8 V & low S/N \\
F187& 17 45 50.542 &  -28 49 23.30 &  19.22     &  18.10     &   17.05    & 14.96 & - & - &  \\
F188& 17 45 50.936 &  -28 49 21.80 &  19.02     &  17.93     &   16.91    & 14.97 & 2(2)  &  $\geq$O8 V & low S/N \\
F189& 17 45 50.595 &  -28 49 23.52 &  18.91     &  17.79     &   16.69    & 14.77 & 3(3) & -  & featureless \\
F190& 17 45 49.759 &  -28 49 19.76 &  19.44     &  18.25     &   17.14    & 14.99 & - & - &  \\
F191& 17 45 51.140 &  -28 49 18.17 &  18.83     &  17.79     &   16.84    & 15.00 & - & - &  \\
F192& 17 45 50.048 &  -28 49 04.94 &  19.70     &  18.43     &   17.16    & 15.00 & - & - &  \\
F193& 17 45 50.383 &  -28 49 09.56 &  19.08     &  17.98     &   16.93    & 15.00 & - & - &  \\
F194& 17 45 50.531 &  -28 49 07.93 &  20.41     &  19.00     &   17.62    & 15.01 & - & - &  \\
F195& 17 45 50.530 &  -28 49 15.30 &  19.11     &  18.00     &   17.01    & 15.03 & - & - &  \\
F196& 17 45 50.463 &  -28 49 07.74 &  19.55     &  18.35     &   17.18    & 15.03 & - & - &  \\
\end{longtable}
{Column  1 indicates the nomenclature for cluster members adopted  by Fi02 and Blum et al. (\cite{blum}), columns 2 and 3 the J2000 co-ordinates, columns 4-6 the 
new HST WFC3 photometry described in Sect. 2.2  
and column 7  F205W filter photometry from Fi02. Column 8 presents the  total number of VLT/SINFONI data-cubes available 
for individual objects, with the number in parentheses being the number of epochs on which these data were obtained ($^{a,b,c}$ 
denotes 
that, respectively, 1,2 or 3  epochs of spectroscopy  were not employed due to low S/N). Column 9 provides a spectral 
classification where 
spectra are available, while the final column provides additional notes including  the presence of radio  (data from Lang et al. 
\cite{lang}, where $^1$ denotes a radio variable source) and X-ray detections (Wang et al. \cite{wang}). As described in Sect. 3, F11, 
F46, F51 and F99 appear to be foreground M stars and so are excluded from this compilation; it would appear likely that other 
interlopers may also be present amongst those stars without current spectral classifications.
}}

\begin{figure*}
\includegraphics[width=14cm,angle=0]{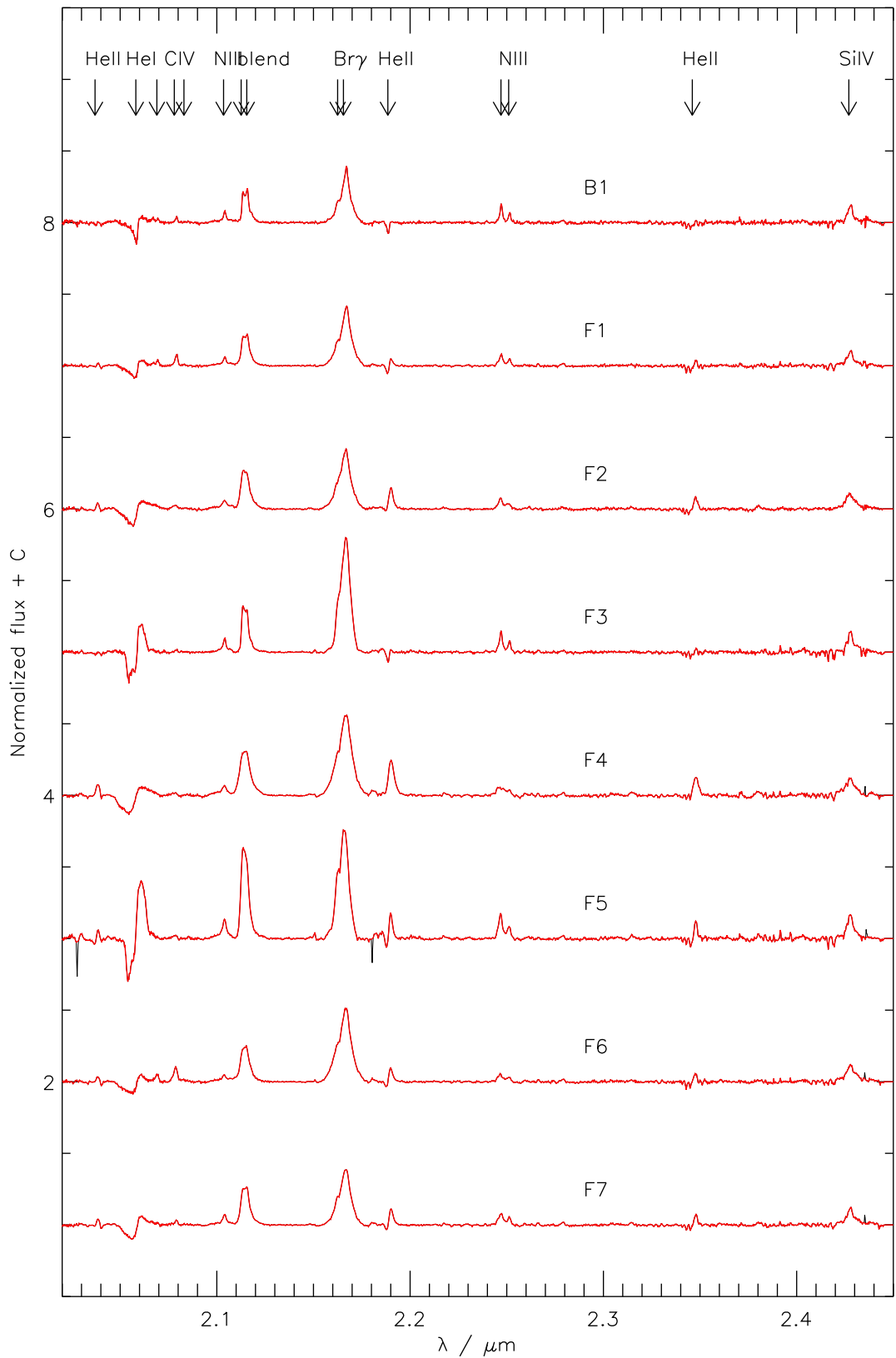}
\caption{Montage of the spectra of all Arches cluster members extracted in this study. Note that spectra 
corresponding to late spectral type interlopers have been excluded. Black lines correspont to unmodified spectra, red lines reflect spectra that have been manually corrected for spurious features.}
\end{figure*}

\begin{figure*}
\includegraphics[width=14cm,angle=0]{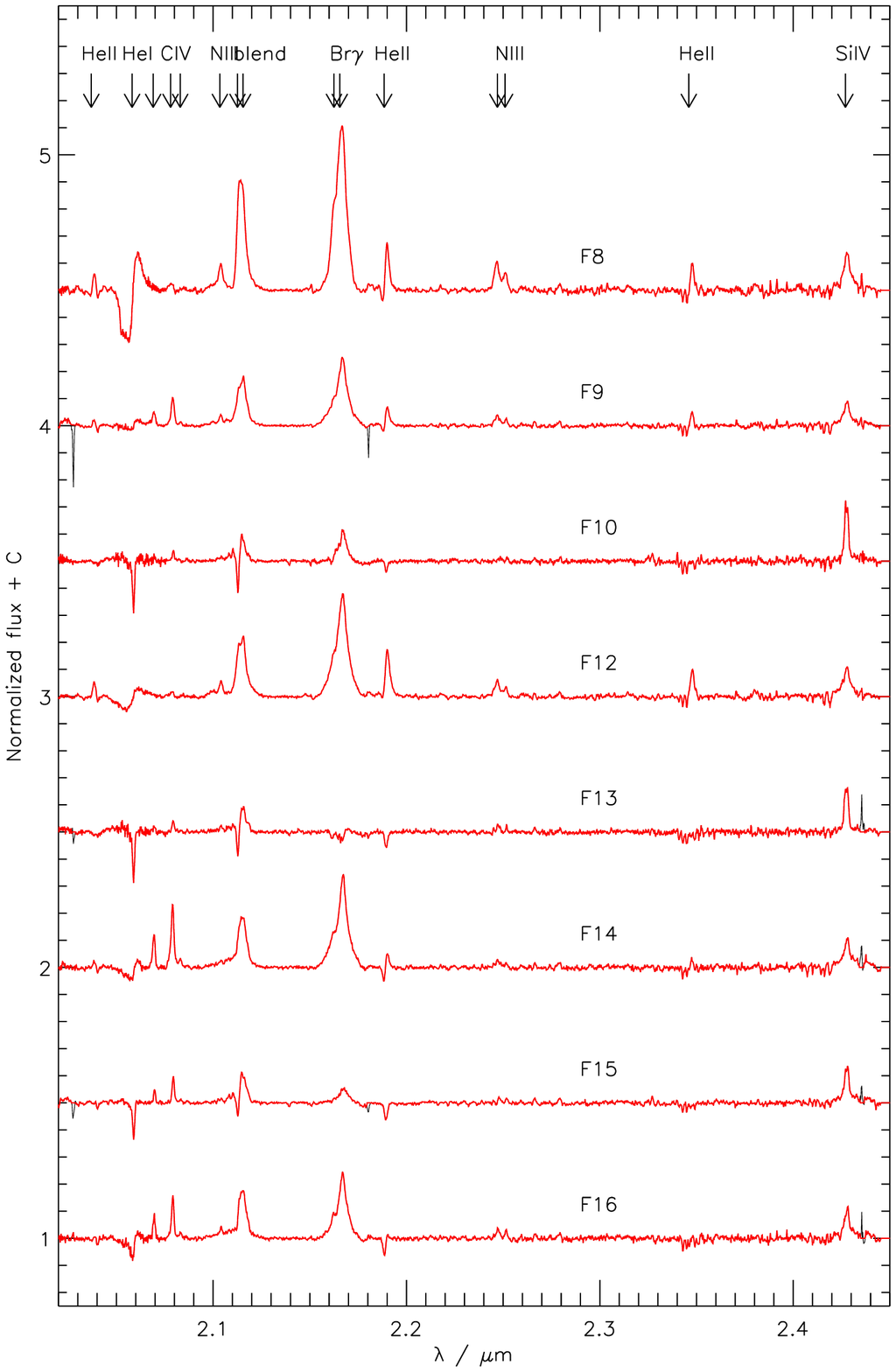}
\end{figure*}

\begin{figure*}
\includegraphics[width=14cm,angle=0]{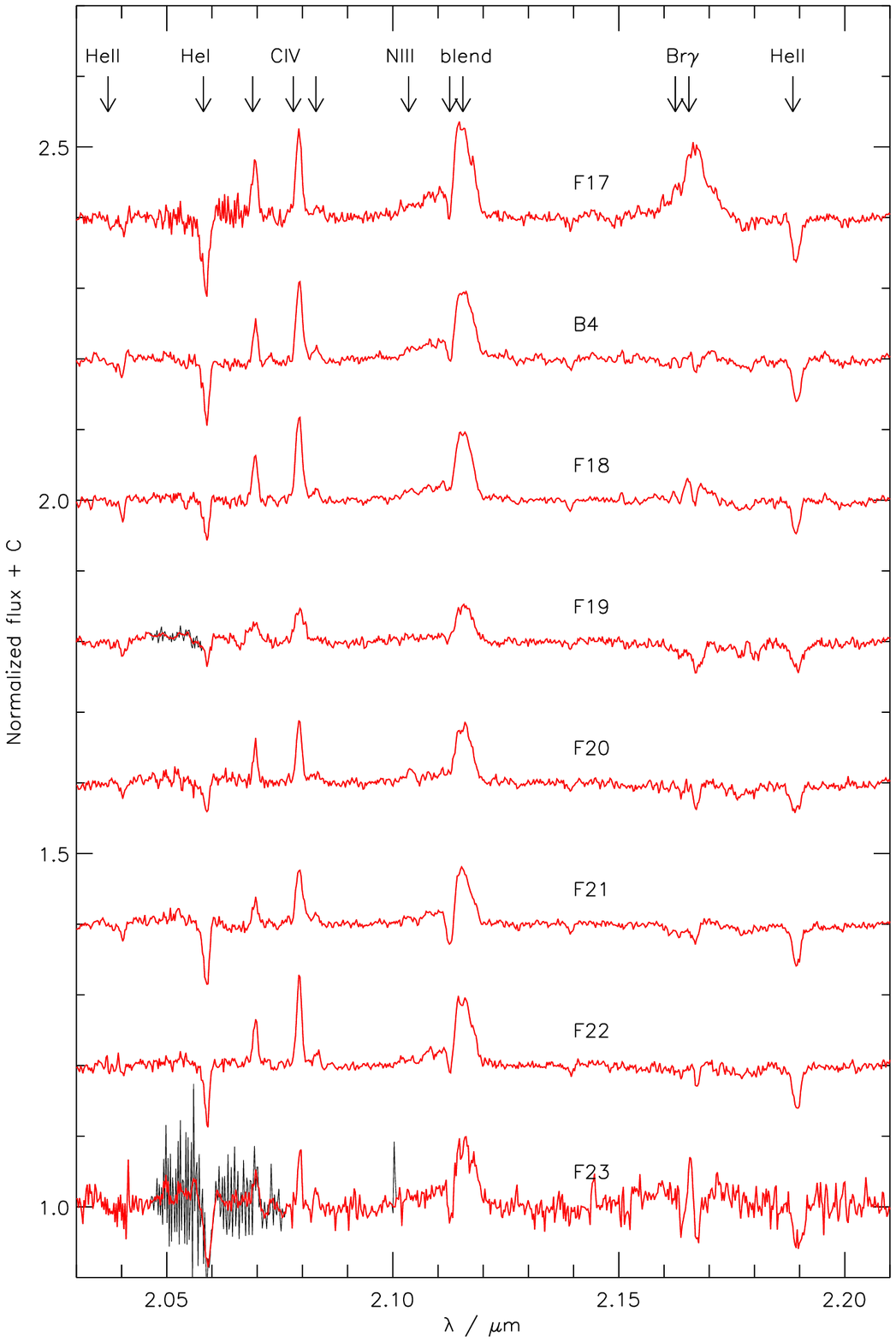}
\end{figure*}

\begin{figure*}
\includegraphics[width=14cm,angle=0]{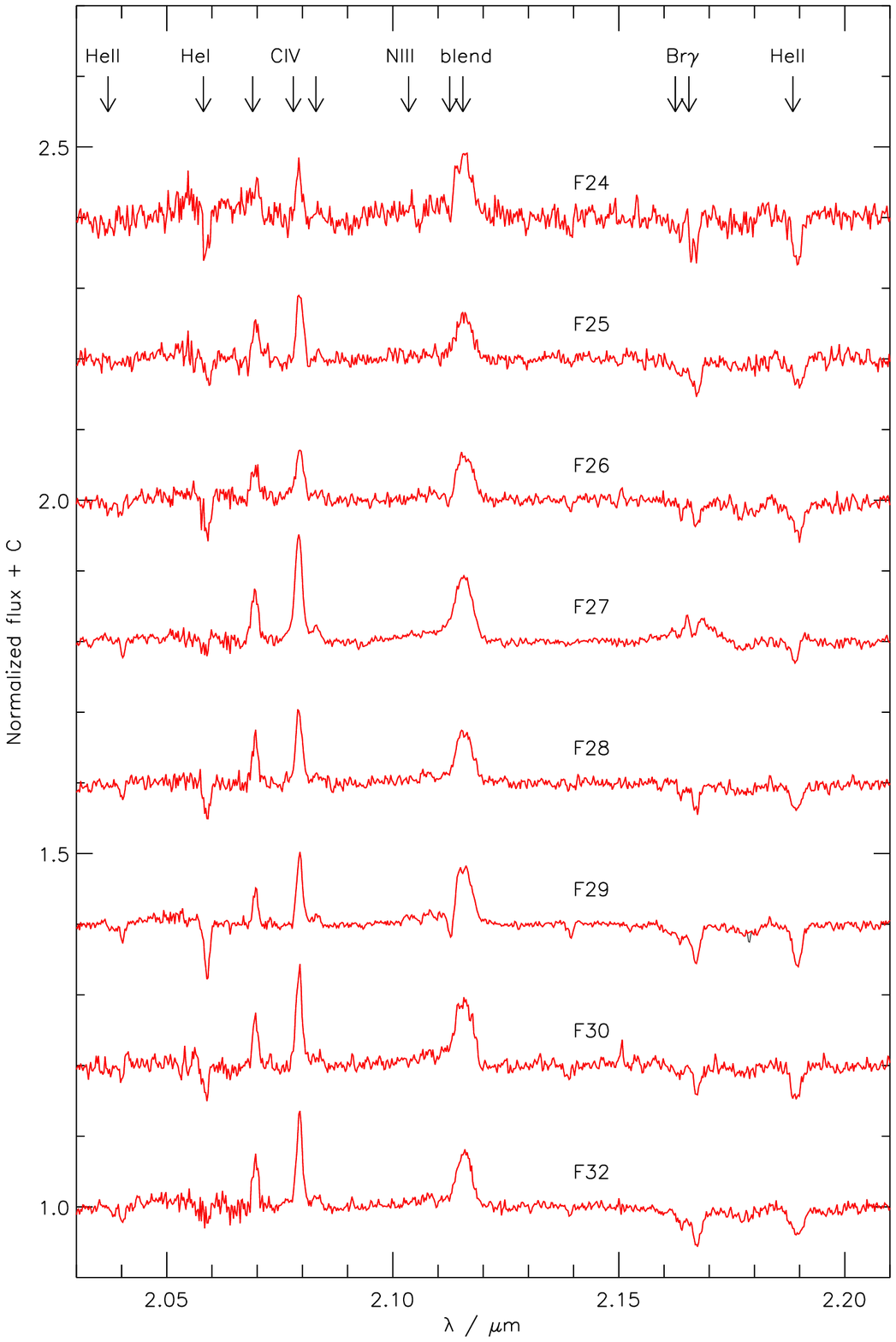}
\end{figure*}

\begin{figure*}
\includegraphics[width=14cm,angle=0]{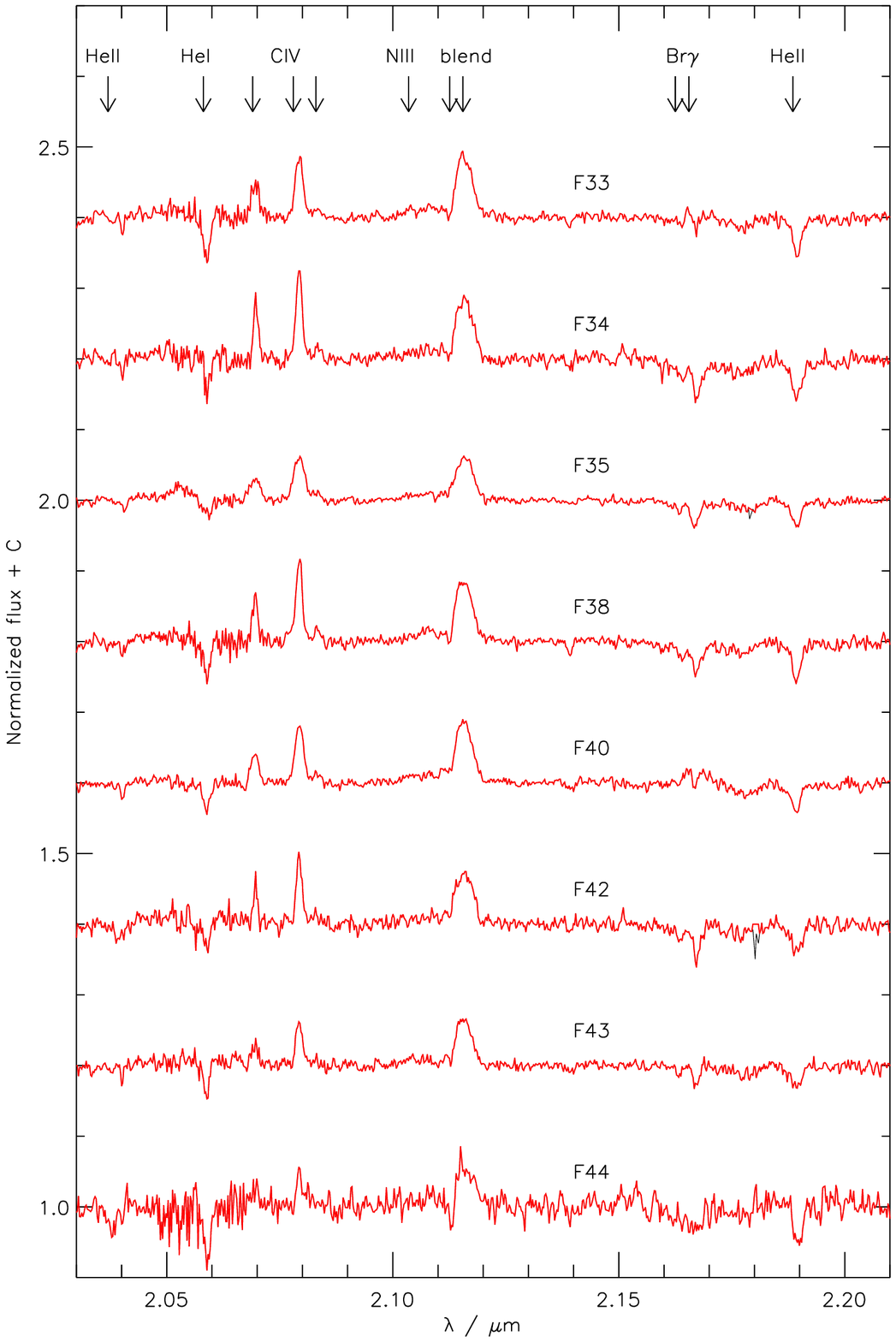}
\end{figure*}

\begin{figure*}
\includegraphics[width=14cm,angle=0]{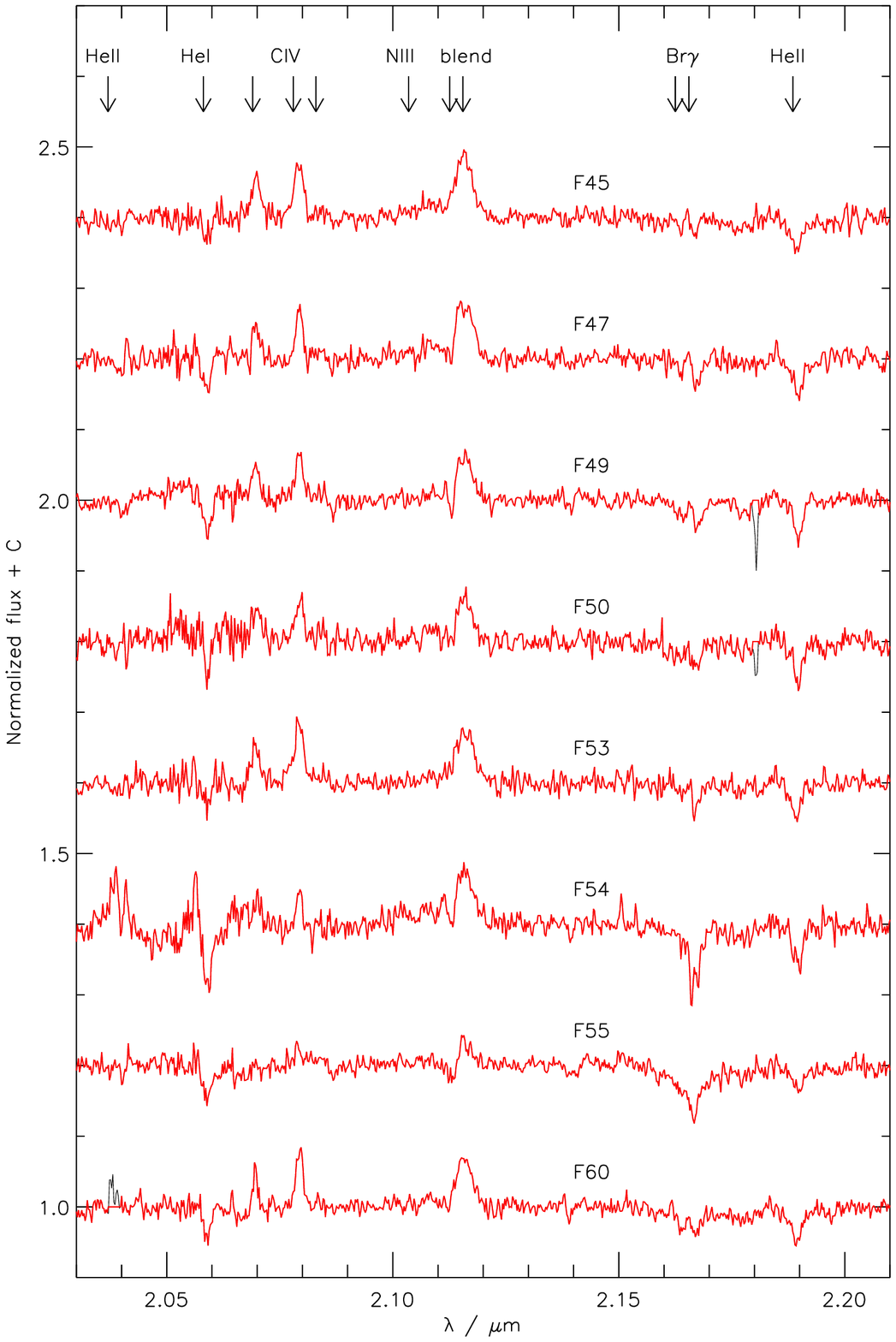}
\end{figure*}

\begin{figure*}
\includegraphics[width=14cm,angle=0]{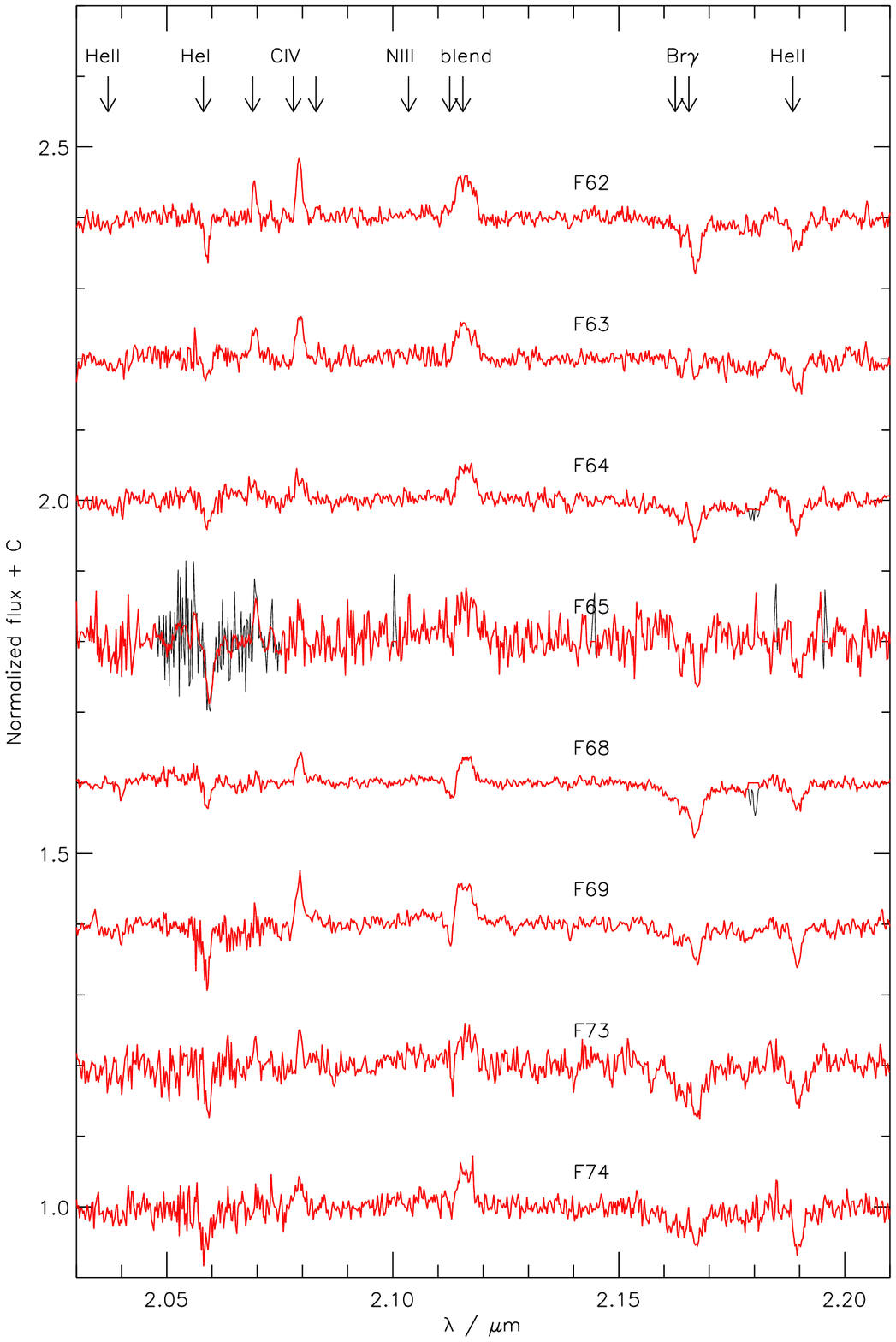}
\end{figure*}

\begin{figure*}
\includegraphics[width=14cm,angle=0]{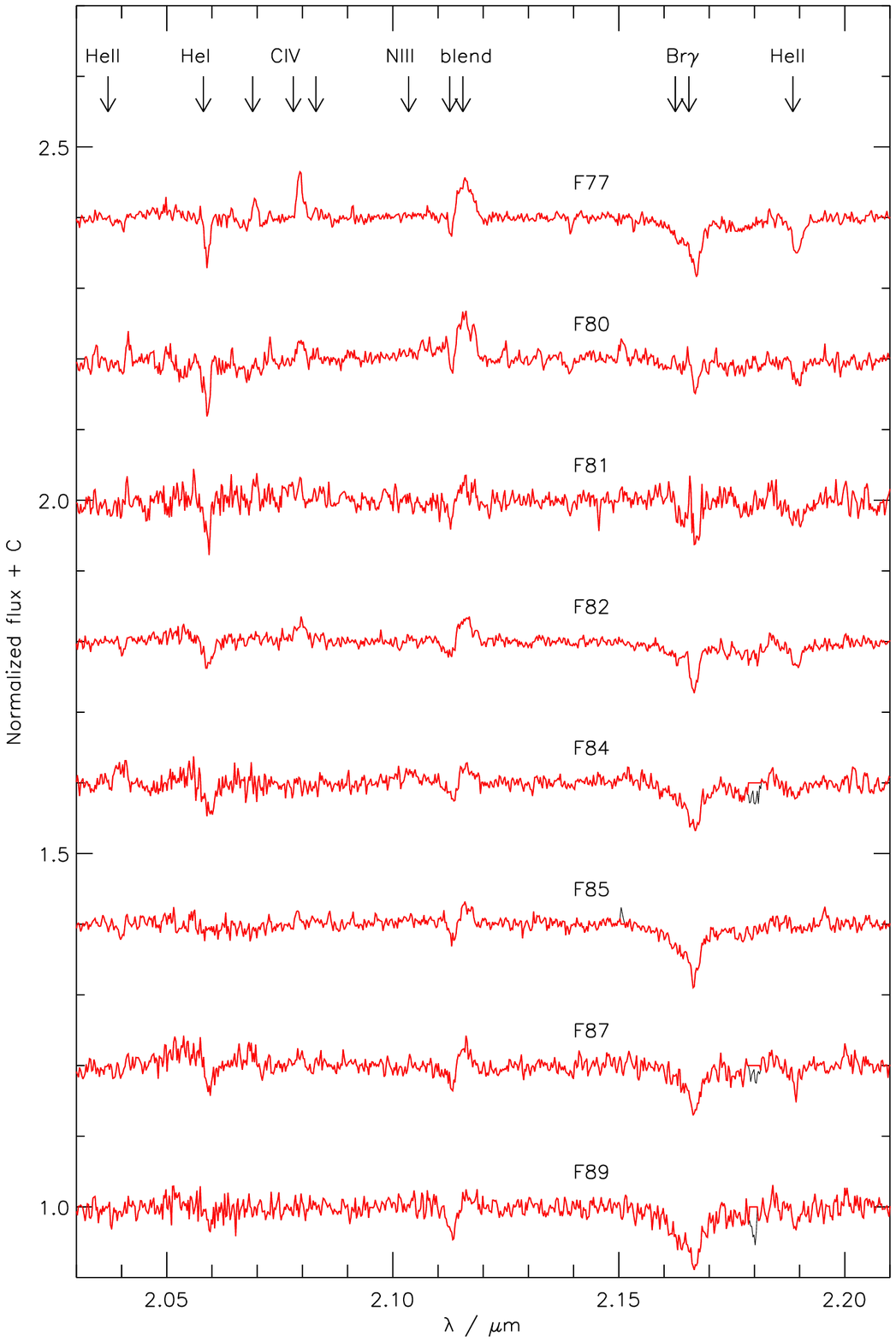}
\end{figure*}

\begin{figure*}
\includegraphics[width=14cm,angle=0]{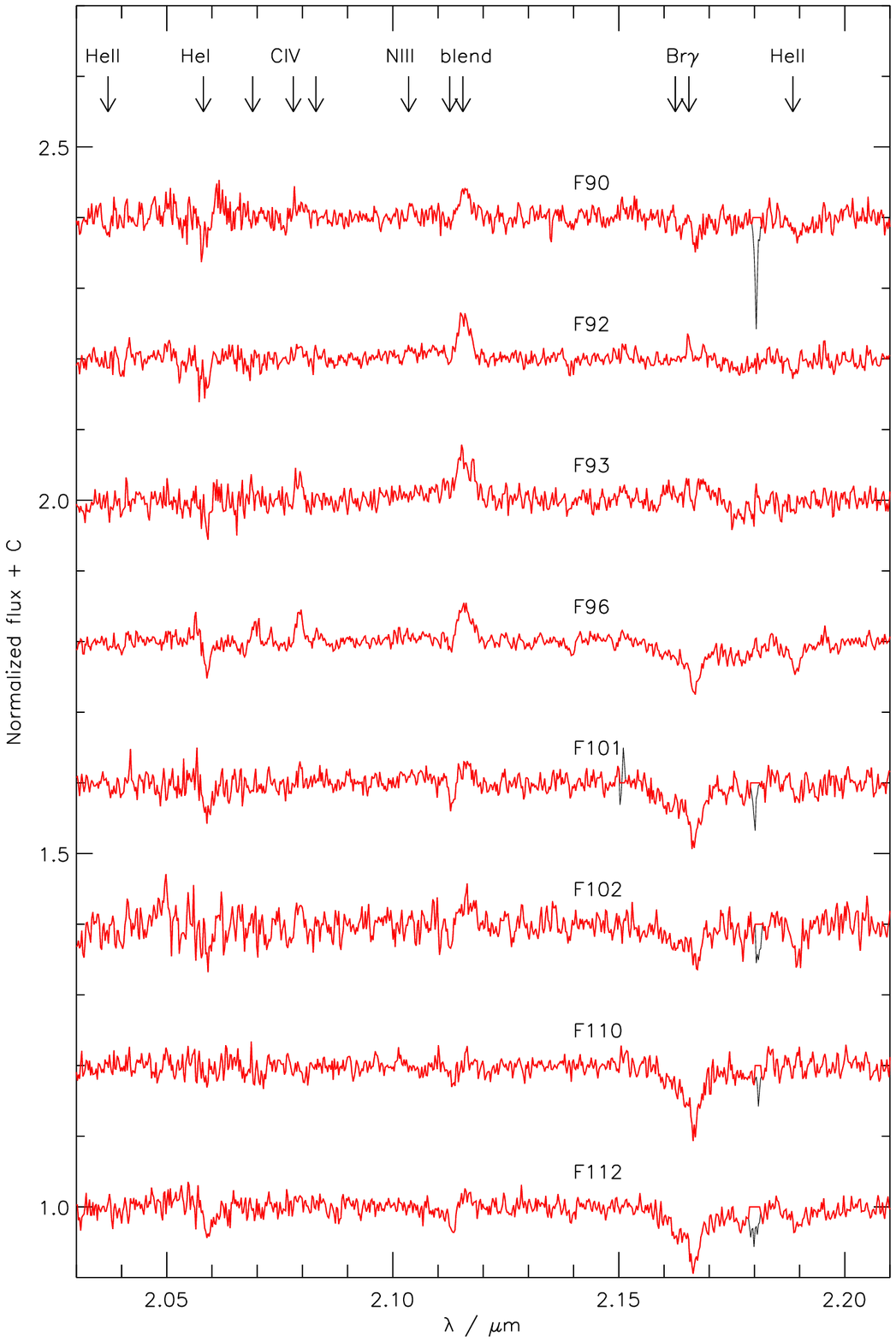}
\end{figure*}

\begin{figure*}
\includegraphics[width=14cm,angle=0]{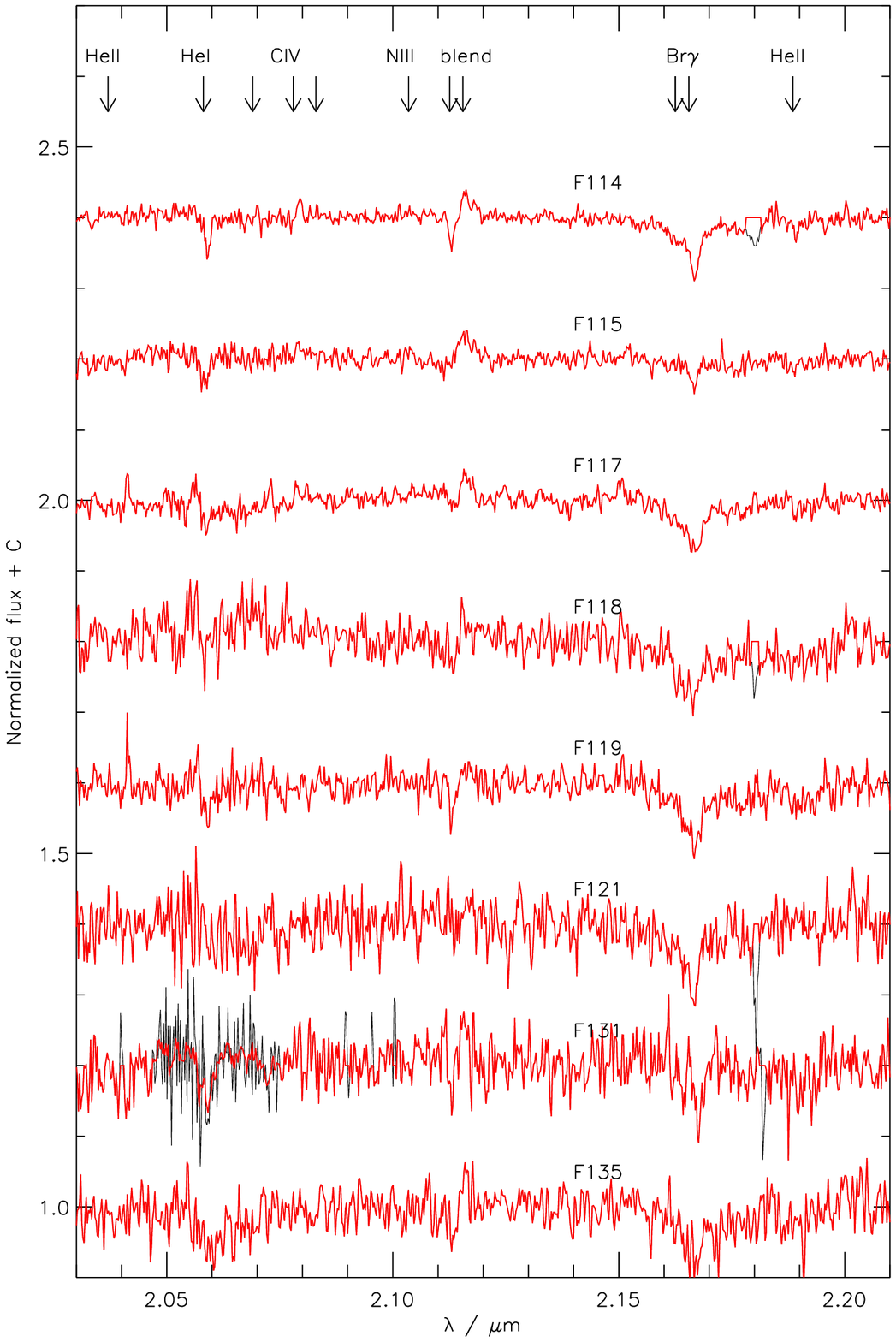}
\end{figure*}

\begin{figure*}
\includegraphics[width=14cm,angle=0]{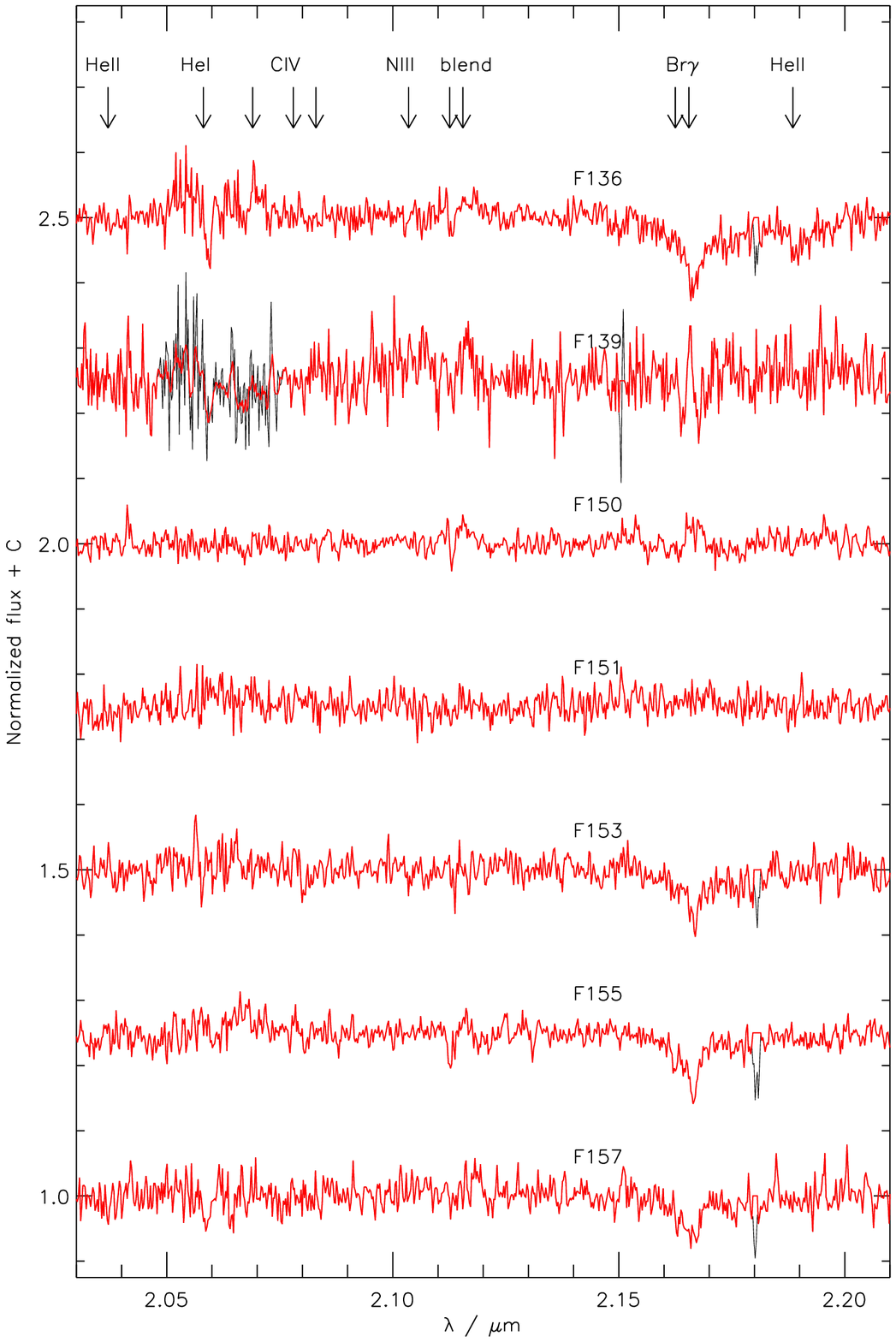}
\end{figure*}

\begin{figure*}
\includegraphics[width=14cm,angle=0]{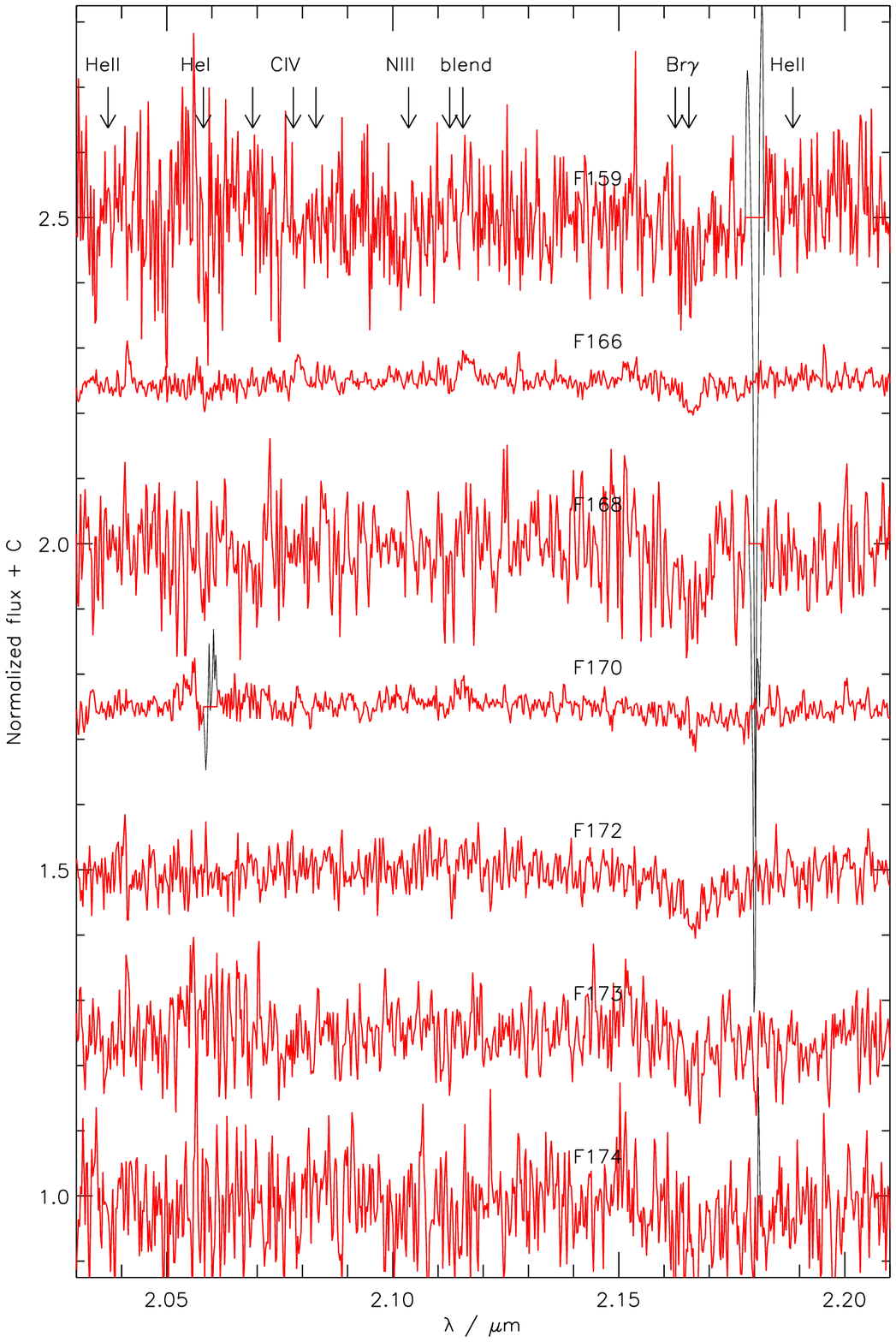}
\end{figure*}

\begin{figure*}
\includegraphics[width=14cm,angle=0]{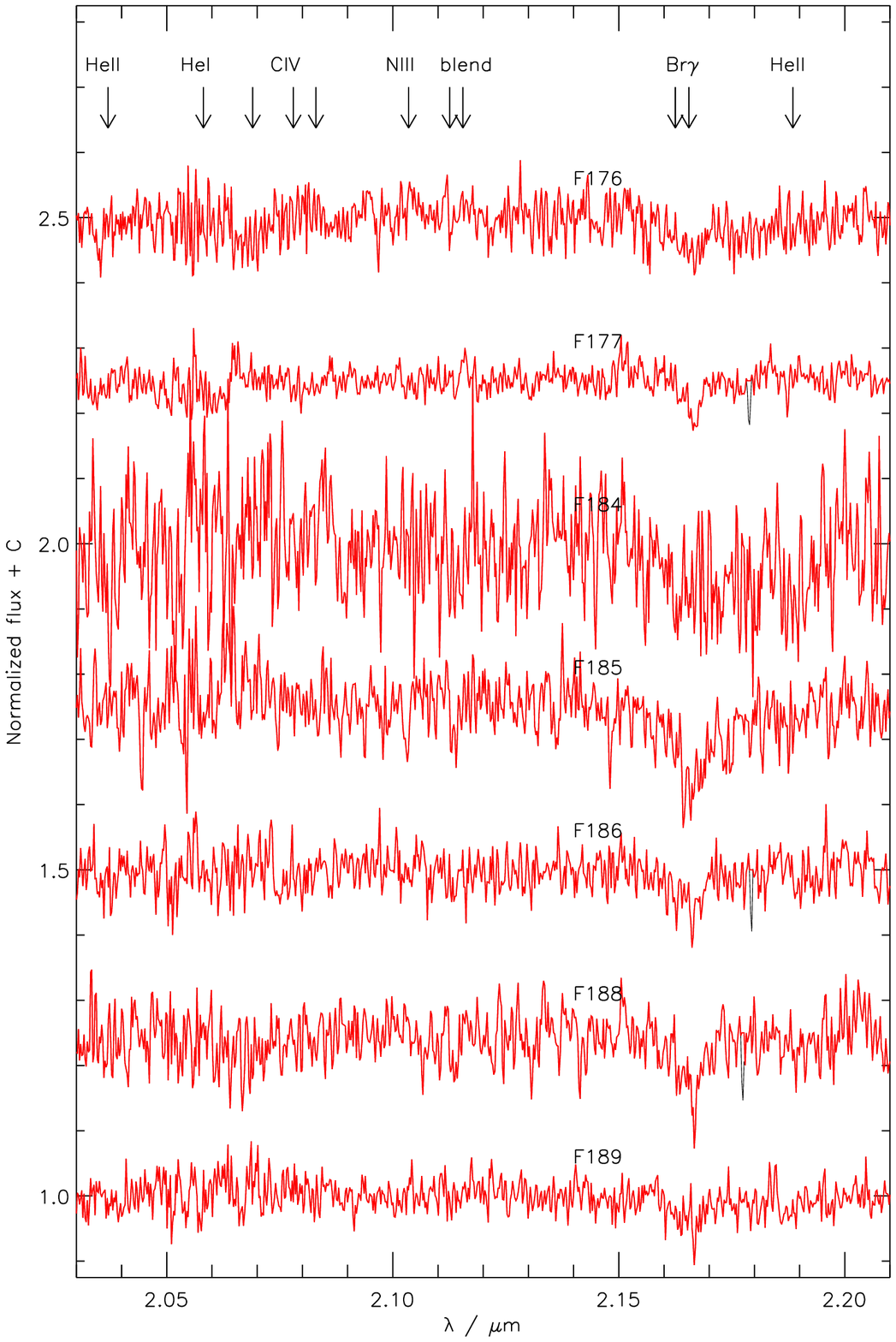}
\end{figure*}

\begin{figure*}
\includegraphics[width=14cm,angle=0]{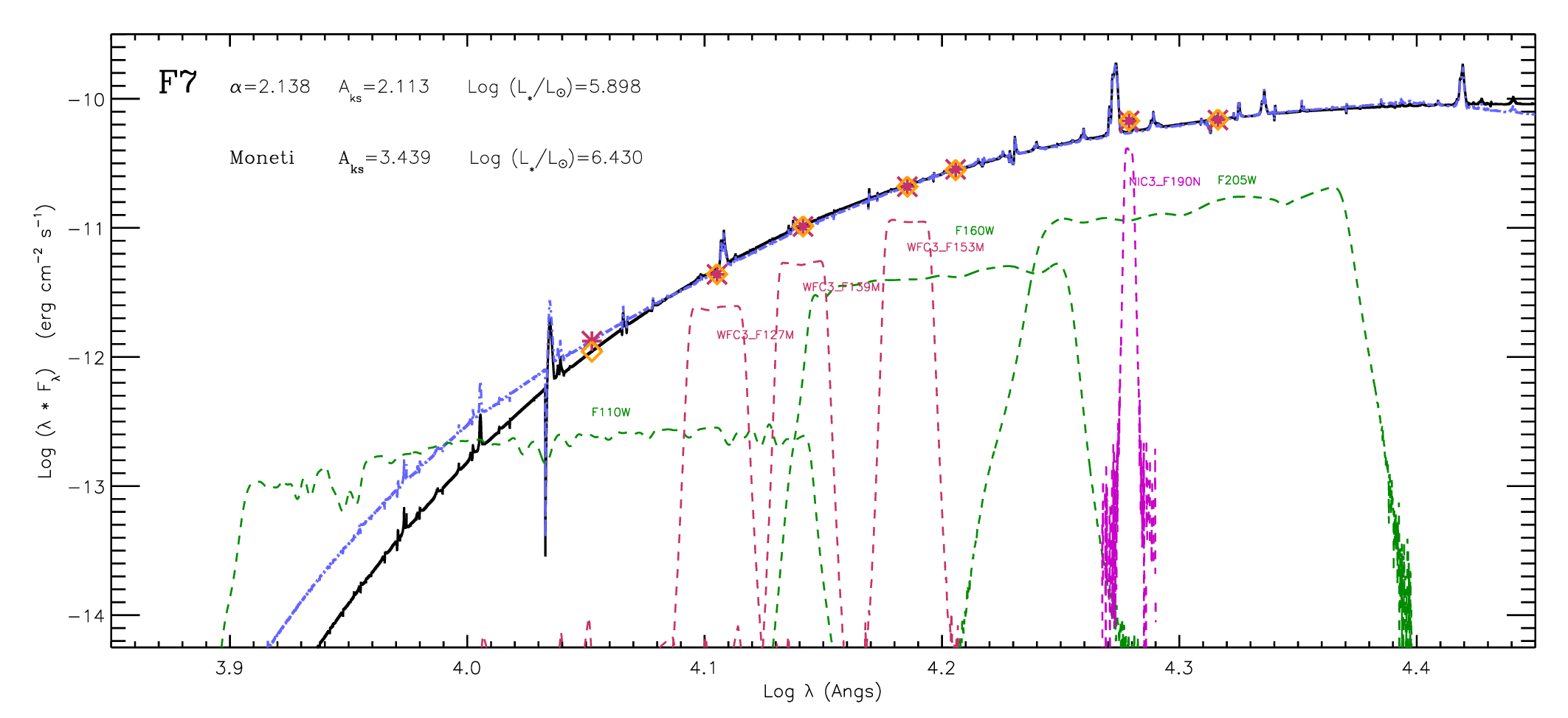}
\includegraphics[width=14cm,angle=0]{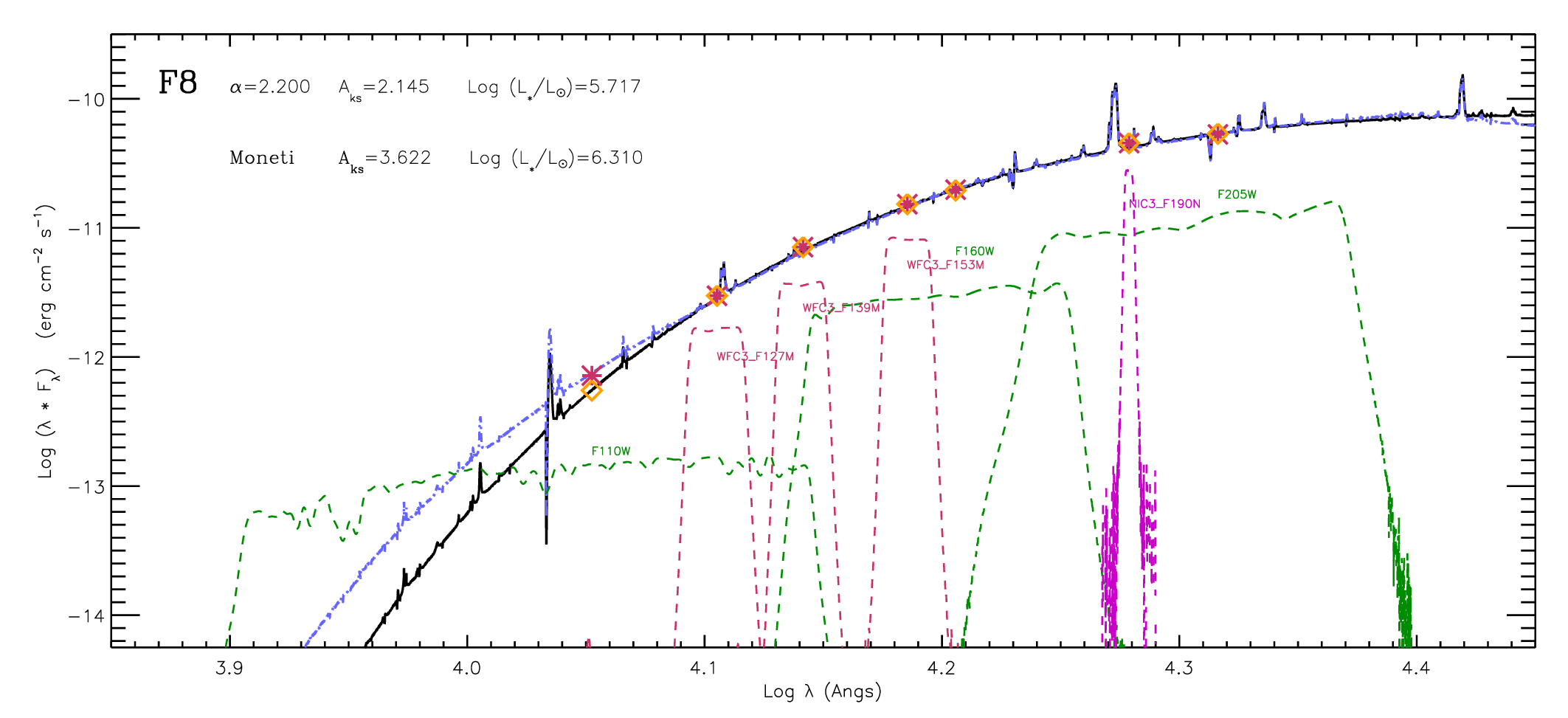}
\includegraphics[width=14cm,angle=0]{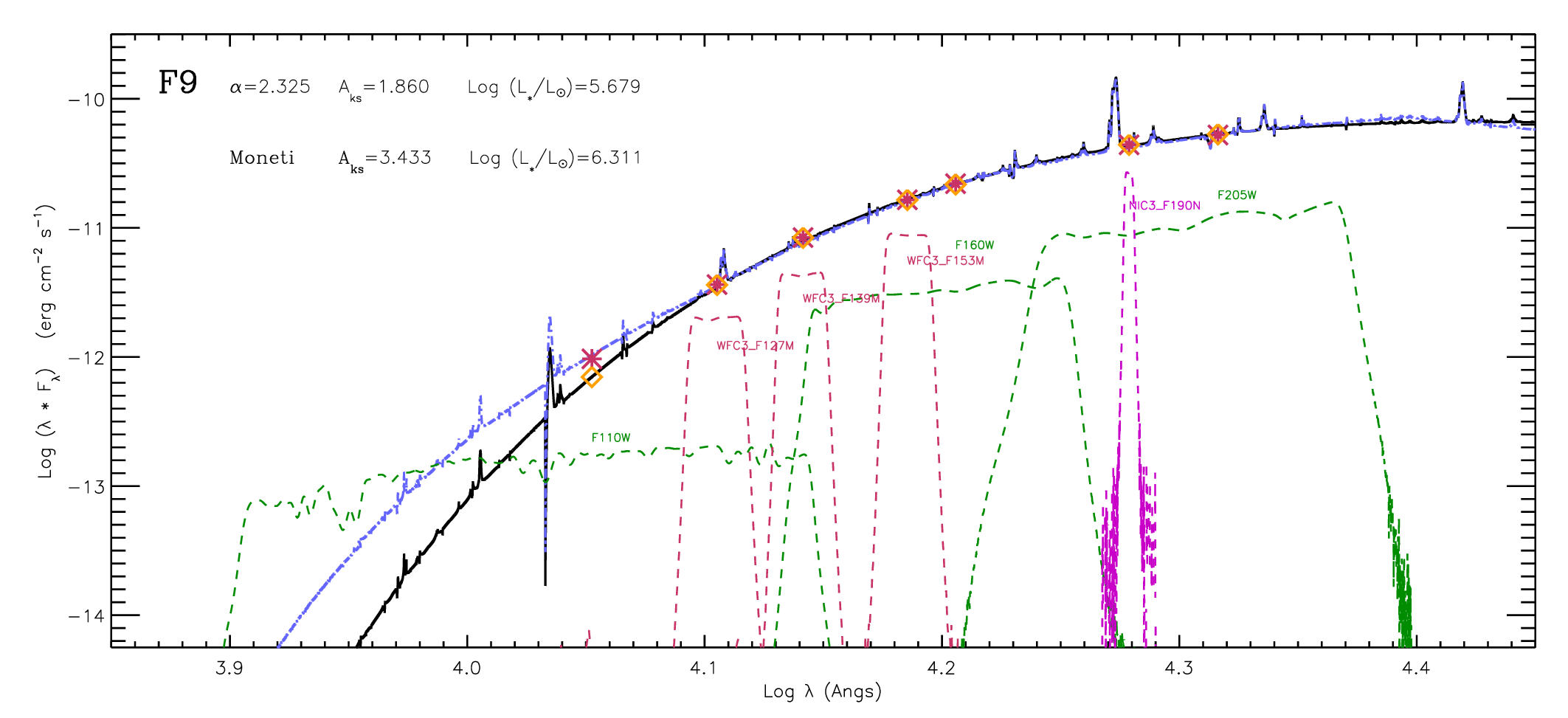}
\caption{Following Fig. 9, synthetic model-atmosphere spectra for the WNLh stars F7, F8 and F9,  computed for two 
differing assumed
interstellar reddening laws, illustrating the dramatic dependence of bolometric luminosity on this choice.
Photometry employed from Table 1, Fi02 and Dong et al. (\cite{dong11}).}
\end{figure*}


\begin{thebibliography}{}
\bibitem[2016]{armard}
Armaed, L., Palacios, A., Charbonnel, C., Gallet, F. \& Bouvier, J. 2016, A\&A, 587, A105

\bibitem[2001]{blum}
Blum, R. D., Schaerer, D., Pasquali, A. et al. 2001, AJ, 122, 1875

\bibitem[2004]{bonanos}
Bonanos, A. Z., Stanek, K. Z., Udalski, A. et al. 2004, ApJ, 611, L33

\bibitem[2001]{bonnell}
Bonnell, I. A., Bate, M. R., Clarke, C. J. \& Pringle, J. E. 2001, MNRAS, 
323, 785

\bibitem[2004]{bonnet}
Bonnet, H., Abuter, R., Baker, A., et al. 2004, The Messenger, 117, 17

\bibitem[1989]{cardelli}
Cardelli, J. A., Clayton, G. C. \& Mathis, J. S. 1989, ApJ, 345, 245

\bibitem[2002]{clark02}
Clark, J. S., Goodwin, S. P., Crowther, P. A., et al. 2002, A\&A, 392, 909


\bibitem[2005]{clark05}
Clark, J. S., Negueruela, I., Crowther, P. A. \& Goodwin, S. P. 2005, A\&A, 434, 949

\bibitem[2011]{clark11}
Clark, J. S., Ritchie, B. W., Negueruela, I. et al. 2011, 531, A28

\bibitem[2013a]{clark13a}
Clark, J. S., Bartlett, E. S., Coe, M. J. et al. 2013a, A\&A, 560, A10


\bibitem[2013b]{clark13}
Clark, J. S., Ritchie, B. W. \&  Negueruela, I. 2013b, A\&A, 560, A11

\bibitem[2014]{clark14}
Clark, J. S., Ritchie, B. W., Najarro, F., Langer, N. \& Negueruela, I.
2014, A\&A, 567, A73

\bibitem[2012]{clarkson}
Clarkson, W. I., Ghez, A. M., Morris, M. R. et al. 2012, ApJ, 751, 132


\bibitem[1996]{cotera}
Cotera, A. S., Erickson, E. F., Colgan, S. W. J. et al.  1996, ApJ, 461, 750

\bibitem[2006]{crowther06}
Crowther, P. A., Hadfield, L. J., Clark, J. S., Negueruela, I. \& Vacca, W.
D. 2006, MNRAS, 372, 1407

\bibitem[2008]{crowther08}
Crowther, P. A. \& Furness, J. P. 2008, A\&A, 492, 111

\bibitem[2010]{crowther10}
Crowther, P. A., Schnurr, O., Hirschi, R. et al. 
2010, MNRAS, 408, 731

\bibitem[2011]{crowther11}
Crowther, P. A. \& Walborn, N. R. 2011, MNRAS, 416, 1311

\bibitem[2016]{crowther16}
Crowther, P. A., Caballero-Nieves, S. M., Bostroem, K. A., et al. 
2016, MNRAS, 458, 624


\bibitem[2012]{davies}
Davies, B., Clark, J. S., Trombley, C. et al. 2012, MNRAS, 419, 1871

\bibitem[2009]{decressin}
Decressin, T., Mathis, S., Palacios, A. et al. 2009, A\&A, 495, 271

\bibitem[2014]{demink}
de Mink, S. E., Sana, H., Langer, N., Izzard, R. G. \& Schneider, F. R N.
2014, ApJ, 782, 7

\bibitem[2000]{dolphin}
Dolphin, A. E., 2000, PASP, 112, 1383


\bibitem[2011]{dong11}
Dong, H., Wang, Q. D., Cotera, A. et al. 2011, MNRAS, 417, 114

\bibitem[2017]{dong17}
Dong, H., Sch\"{o}del, R., Williams, B. F. et al. 2017, MNRAS, 470, 3427

\bibitem[1995]{drissen}
Drissen, L., Moffat, A. F. J., Walborn, N. R. \&  Shara, M. M. 1995, AJ, 110, 2235

\bibitem[2003]{eisenhauer}
Eisenhauer, F., Abuter, R., Bickert, K., et al. 2003, in Society of Photo-Optical
Instrumentation Engineers (SPIE) Conference Series, Vol. 4841, Instrument
Design and Performance for Optical/Infrared Ground-based Telescopes, ed.
M. Iye \& A. F. M. Moorwood, 1548–1561


\bibitem[2009]{espinoza}
Espinoza, P., Selman, F. J. \& Melnick, J. 2009, A\&A, 501, 563

\bibitem[2017]{fenech}
Fenech, D. M., Clark, J. S., Prinja, R. K. et al. 2017, MNRAS, 464, L75

\bibitem[1999a]{figer99a}
Figer, D. F., McLean, I. A. \& Morris, M. 1999a, ApJ, 514, 202

\bibitem[1999b]{figer99b}
Figer, D. F., Kim, S. S., Morris, M., et al. 1999b, ApJ, 525, 750



\bibitem[2002]{figer02}
Figer, D. F., Najarro, F., Gilmore, D. et al. 2002, ApJ, 581, 258 (Fi02)

\bibitem[1995]{gehrz95}
Gehrz, R. D., Hayward, T.L., Houck, J. R. et al. 1995, ApJ, 439, 417

\bibitem[2001]{gehrz01}
Gehrz, R. D., Smith, N., Jones, B., Puetter, R. \& Yahil, A. 2001, ApJ,559, 395

\bibitem[2003]{gies}
Gies, D. R. 2003, in A Massive Star Odyssey: From Main Sequence to Supernova, 
ed. K. van der Hucht, A. Herrero, \& C. Esteban, IAU Symp., 212, 91


\bibitem[2013]{groh13}
Groh, J. H., Meynet, G., Georgy, C.  \& Ekstr\"{o}m, S. 2013, A\&A, 558, 
A131


\bibitem[2014]{groh14}
Groh, J. H., Meynet, G., Ekstr\'{o}m, S. \&  Georgy, C. 2014, A\&A, 564, 
A30

\bibitem[2007]{grundstrom}
Grundstrom, E. D., Gies, D. R., Hellwis, T. C. et al. 2007, ApJ, 667, 505

\bibitem[2013]{habibi}
Habibi, M., Stolte, A., Brandner, W. Hu{\ss}man, B. \& Motohara, K. 
2013, A\&A, 556, A26


\bibitem[1996]{hanson96}
Hanson, M. M., Conti, P. S. \& Rieke, M. J. 1996, ApJS, 107, 281

\bibitem[2005]{hanson05}
Hanson, M. M., Kudritzki, R.-P., Kenworthy, M. A., Puls, J. \&
Tokunaga, A. T. 2005, ApJS, 161, 154 

\bibitem[1998]{hillier98}
Hillier, D. J. \& Miller, D. L. 1998, ApJ, 496, 407

\bibitem[1999]{hillier99}
Hillier, D. J. \& Miller, D. L. 1999, ApJ, 519, 354

\bibitem[1986]{horne}
Horne, K. 1986, PASP,98, 609

\bibitem[2010]{kastner}
Kastner, J. H., Buchanan, C., Sahai, R., Forrest, W. J. \& Sargent, B. A., 2010, AJ, 139, 1993


\bibitem[2006]{kim}
Kim, S. S., Figer, D. F., Kudritzki, R. P. \& Najarro, F.
2006, ApJ, 653, L113

\bibitem[2005]{lang}
Lang, C. C., Johnson, K. E., Goss, W. M. \&  Rodr\'{i}guez, L. F 2005, AJ, 130,
2185


\bibitem[2009]{liermann09}
Liermann, A., Hamann, W.-R. \&  Oskinova, L. M. 2009, A\&A, 494, 1137

\bibitem[2009]{liermann12}
Liermann, A., Hamann, W.-R. \&  Oskinova, L. M. 2014, 540, A14

\bibitem[2009]{linder}
Linder, N., Rauw, G., Manford, J. et al. 2009, A\&A, 495, 231


\bibitem[submitted]{lohr16a}
Lohr, M. E., Clark, J. S., Najarro, F, et al. A\&A, submitted (Paper II)

\bibitem[in prep.]{lohr16b}
Lohr, M. E., Clark, J. S., Patrick, L. et al.  A\&A, in prep. (Paper III)


\bibitem[2014]{lorenzo14}
Lorenzo, J., Negueruela, I., Baker, A. F. K. Val, et al. 2014, A\&A, 572, A110

\bibitem[1999]{lutz}
Lutz, D. 1999, The Universe as Seen by ISO, ed. P. Cox \&
M. F. Kessler, ESA-SP 427, 623


\bibitem[2012]{mahy}
Mahy, L., Gosset, E., Sana, H. et al. 2012, A\&A, 540, A97




\bibitem[2015]{MA}
Ma{\'i}z-Apell{\'a}niz, J., Negueruela, I., Barb{\'a}, R. H. et al. 2015, 
A\&A, 579, A108


\bibitem[2005]{martins05}
Martins, F., Schaerer, D. \& Hillier, D. J. 2005, A\&A, 436, 1049

\bibitem[2006]{martins06}
Martins, F. \& Plez, B. 2006, A\&A, 457, 637


\bibitem[2007]{martins07}
Martins, F., Genzel, R., Hillier, D. J., et al. 2007 A\&A, 468, 233


\bibitem[2008]{martins08}
Martins, F., Hillier, D. J., Paumard, T. et al. 2008, A\&A, 478, 219

\bibitem[2017]{martins16}
Martins, F. \& Palacios, A. 2017, A\&A, 598, A56

\bibitem[2001]{massey}
Massey, P., DeGioia-Eastwood, K. \& Waterhouse, E. 2001, AJ, 121, 1050

\bibitem[2010a]{mauerhan10a}
Mauerhan, J. C., Muno, M. P., Morris, M. R., Stolovy, S. R. \&
Cotera, A. 2010, ApJ, 710, 706

\bibitem[2010b]{mauerhan10b}
Mauerhan, J. C., Cotera, A., Dong H. et al. 2010b, ApJ, 725, 188

\bibitem[2015]{mauerhan15}
Mauerhan, J. C., Smith, N., Van Dyk, S. D. et al. 2015, MNRAS, 450, 2551

\bibitem[2008]{mayer}
Mayer, P., Harmanec, P., Nesslinger, S., et al. 2008, A\&A, 481, 183


\bibitem[2008]{melena}
Melena, N. W., Massey, P., Morrell, N. I. \& Zangari, A. M. 2008, AJ, 135, 878

\bibitem[2003]{meynet}
Meynet, G., \& Maeder, A. 2003, A\&A, 404, 975

\bibitem[2001]{moneti}
Moneti, A., Stolovy, S., Blommaert, J. A. D. L., Figer, D. F., Najarro, F. 2001, A\&A, 366, 106

\bibitem[1995]{nagata}
Nagata, T., Woodward, C. E., Shure, M. \& Kobayashi, N. 1995, AJ, 109, 1676


\bibitem[2004]{najarro}
Najarro, F., Figer, D. F., Hillier, D. J. \&  Kudritzki, R. P. 2004, ApJ, 611, L105


\bibitem[2008]{iggy08}
Negueruela, I., Marco, A., Herrero, A. \& Clark, J. S. 2008, A\&A, 487, 575


\bibitem[2010]{iggy10}
Negueruela, I., Clark, J. S. \& Ritchie, B. W. 2010, A\&A, 516, A78

\bibitem[2009]{nishiyama}
Nishiyama, S., Tamura, M., Hatano, H. et al. 2009, ApJ, 696, 1407

\bibitem[1985]{rieke}
Rieke, G. H. \& Lebofsky, M. J. 1985, ApJ, 288, 618
 
\bibitem[2010]{ritchie}
Ritchie, B. W., Clark, J. S., Negueruela, I. \& Langer, N. 2010, A\&A, 516, A78

\bibitem[2018]{rosslowe}
Rosslowe, C. K. \& Crowther, P. A. 2018, MNRAS, 473, 2853

\bibitem[2012]{sana12}
Sana, H., de Mink, S. E., de Koter, A., et al., 2012, Science, 337, 444

\bibitem[2014]{schneider14}
Schneider, F. R. N., Izzard, R. G., de Mink, S. E., et al. 2014, ApJ, 780, 117

\bibitem[2015]{schneider15}
Schneider, F. R. N., Izzard, R. G., Langer, N. \& de Mink, S. E. 2015, ApJ, 805, 20

\bibitem[2008]{schnurr}
Schnurr, O., Casoli, J., Chen{\'e}, A.-N., Moffat, A. F. J. \& 
St-Louis, N. 2008, MNRAS, 389, L38

\bibitem[2010]{schoedel}
Schoedel, R., Najarro, F., Muzic, K. \& Eckart, A. 2010, A\&A, 511, A18

\bibitem[1999]{sch}
Schweikhardt, J., Scmutz, W., Stahl, O., Szeifert, Th. \& Wolf, B., 1999,
A\&A, 347, 127

\bibitem[1998]{serabyn}
Serabyn, E., Shupe, D. \& Figer, D. F. 1998, Nature, 394, 448

\bibitem[1987]{shu}
Shu, F. H., Adams, F. C. \& Lizano, S. 1987, ARA\&A, 25, 23


\bibitem[2006]{smith06}
Smith, N. 2006, MNRAS, 367, 763

\bibitem[2008]{smith08}
SMith, N. \& Conti, P. S. 2008, ApJ, 679, 1467


\bibitem[2002]{stolte02}
Stolte, A., Grebel, E. K., Brandner, W. \& Figer, D. F. 2002, A\&A, 394, 
459 

\bibitem[2005]{stolte05}
Stolte, A., Brandnder, W., Grebel, E. K., Lenzen, R. \& Lagrange, A.-M.  2005, ApJ, 628, L113

\bibitem[2010]{stroud}
Stroud, V. E., Clark, J. S., Negueruela, I. et al. 2010, A\&A, 511, A84

\bibitem[2011]{taylor}
Taylor, W. D., Evans, C. J., Sana, H. et al. 2011, A\&A, 530, L10

\bibitem[2016]{tramper}
Tramper, F., Sana, H., Fitzsimons, N. E., et al. 2016, MNRAS, 455, 1275

\bibitem[1998]{van}
van Bever, J. \& Vanbeveren, D. 1998, A\&A, 334, 21

\bibitem[2001]{vazquez}
V\'{a}zquez, R. A. \& Baume, G. 2001, A\&A, 371, 908

\bibitem[2006]{wang}
Wang, Q., D., Dong, H. \& Lang, C. 2006, MNRAS, 371, 38

\bibitem[2010]{weidner}
Weidner, C. \& Vink, J. S. 2010, A\&A, 524, A98

\bibitem[2015]{wright}
Wright, N. J., Drew, J. E. \& Mohr-Smith, M. 2015, MNRAS, 449, 741

\bibitem[2013]{yasarsoy}
Yasarsoy, B. \& Yakut, K. 2013, AJ, 145, 9

\bibitem[2007]{zinnecker}
Zinnecker, H. \& Yorke, H. W. 2007, ARA\&A, 45, 481


\end{thebibliography}
\end{document}